\documentclass[onecollarge,natbib]{svjour2}
\bibpunct{[}{]}{;}{n}{}{,} 
\smartqed  
\usepackage{graphicx}
\usepackage{tabularx}
\usepackage{cases}
\usepackage{multirow}
\usepackage{latexsym}
\journalname{Few-body systems}

\begin{document}

\title{Efimov Resonances in Ultracold Quantum  Gases
}


\author{F. Ferlaino \and A. Zenesini \and M. Berninger  \and  B.~Huang \and H.-C. N\"{a}gerl \and  R. Grimm
}


\institute{F. Ferlaino  \and A. Zenesini \and M. Berninger  \and B. Huang \and H.-C. N\"{a}gerl \at
              Institut f\"ur Experimentalphysik and Zentrum f\"ur Quantenphysik, Universit\"at Innsbruck, 6020 Innsbruck, Austria,
               \email{francesca.ferlaino@uibk.ac.at}           
           \and
           R. Grimm \at
              Institut f\"ur Quantenoptik und Quanteninformation,
 \"Osterreichische Akademie der Wissenschaften, 6020 Innsbruck,
 Austria \&  Institut f\"ur Experimentalphysik and Zentrum f\"ur Quantenphysik, Universit\"at Innsbruck, 6020 Innsbruck, Austria
}

\date{Received: date / Accepted: date}

\maketitle

\begin{abstract}
Ultracold atomic gases have developed into prime systems for experimental studies of Efimov three-body physics and related few-body phenomena, which occur in the universal regime of resonant interactions. In the last few years, many important breakthroughs have been achieved, confirming basic predictions of universal few-body theory and deepening our understanding of such systems. We review the basic ideas along with the fast experimental developments of the field, focussing on ultracold cesium gases as a well-investigated model system. Triatomic Efimov resonances, atom-dimer Efimov resonances, and related four-body resonances are discussed as central observables. We also present some new  observations of such resonances, supporting and complementing the set of available data.
\end{abstract}

\section{Introduction}
	\label{sec:1}
	
In the early seventies, the Russian theoretical physicist V.\,Efimov discovered a fundamental property of quantum three-body systems. He predicted that a system of three identical bosons with resonant pairwise interaction features an infinite series of three-body bound states \cite{Efimov1970ela}. This surprising phenomenon can occur in a resonantly interacting three-body system regardless of the specific nature and the short-range details of the inter-particle forces.
Despite its generality, the Efimov effect remained elusive to observations for a long time. Over the last forty years, experimental realizations of the Efimov effect have been proposed for a number of different systems belonging to nuclear, molecular, and atomic physics, such as triton compounds, halo nuclei, trimers of $^4$He atoms, and  ultracold quantum gases. The variety of these systems represents very different energy regimes, ranging from MeV in nuclear physics to peV for ultracold atomic systems.

After more than 35 years of intense searches an Efimov trimer state was finally observed in summer 2005 in collisional studies on an ultracold gas of cesium atoms \cite{Kraemer2006efe}. Since then, an increasing number of experiments with ultracold gases showed clear evidence of the Efimov effect \cite{Ottenstein2008cso, Huckans2009tbr, Knoop2009ooa, Zaccanti2009ooa, Barontini2009ooh, Gross2009oou, Pollack2009uit, Nakajima2010nea, Gross2010nsi, Lompe2010ads, Nakajima2011moa}, confirming its general character and pointing towards related novel few-body phenomena
\cite{Efimov2009gtt, Greene2010uif, Ferlaino2010fyo}. The main observables in many of these experiments are {\em Efimov resonances} \cite{Efimov1979lep}, which occur when an Efimov state couples to the collision threshold.

Here we review the basic features of Efimov physics in ultracold quantum gases. We discuss the experimental observations with particular emphasis on our results on Cs, obtained in a series of experiments over the past five years \cite{Kraemer2006efe, Knoop2009ooa, Ferlaino2009efu, Berninger2011uot}. Section \ref{sec:3} introduces basic elements of two-body scattering theory and the concept of universality.  Section\,\ref{sec:4} introduces Efimov's scenario and describes its impact on ultracold gases. Section \ref{sec:5main} focuses on the experimental observations of Efimov resonances in ultracold cesium gases, while Section \ref{sec:9} discusses evidence obtained for four-body bound states tied to an Efimov trimer and presents new experimental observations on this subject.
Section \ref{sec:10others} discusses the observations on Efimov physics obtained in a variety of other ultracold systems.

\section{Two-body interaction and universality}
	\label{sec:3}
	
Here we summarize the most important concepts of two-body scattering physics in order to set the stage for the discussion on few-body phenomena. Sec.\,\ref{basic} introduces the scattering length, starting from the general case and then focusing on atomic systems and magnetic Feshbach resonances. In Sec.\,\ref{uni} we introduce the concept of two-body universality. In Sec.\,\ref{frcesium} we describe the particular case of cesium atoms. The reader can refer to Ref.\,\cite{Taylor1983stt} for a complete review on scattering theory and to Ref.\,\cite{Chin2010fri} for Feshbach resonances in ultracold gases.

\subsection{\textit{Basic scattering concepts}}
\label{basic}

\textit{Scattering length} $-$ The two-body scattering length is the essential quantity to characterize elastic collisions between particles at ultralow energies. The Schr\"{o}dinger equation can be solved by superposition of an incoming plane wave and of an outgoing scattered wave. In general the latter contains different partial wave contributions but at very low collisional energies $E \rightarrow 0$, zero orbital angular momentum collisions are predominant ($s$-wave collisions). At long distance the wavefunction of the system approaches its asymptotic form, while a fast oscillation is present in the range of the interaction potential. The phase shift $\phi$ between the incoming and outgoing wave at long distances contains the relevant information on the collision. The phase $\phi$ is generally a function of the momentum $\hbar k$ of the colliding particles but, in the limit of zero energy, it can be expanded in powers of $k^2$ and related to the $s$-wave scattering length $a$ through \cite{Gao1998qdt}
\begin{equation}
k \cot(\phi(k))=-\frac{1}{a}+\frac{r_0 k^2}{2},
\label{phase}
\end{equation}
where $r_0$ is the so-called \textit{effective range}, which is determined from the long-range behavior of the potential.
The term $r_0 k^2/2$ represents the next term in the expansion of the phase shift for $k\neq0$. However, when the particle momentum is much smaller than $\hbar/r_0$, the effect of the interactions is indistinguishable from the one of a contact interaction \cite{Braaten2006uif}.

\begin{figure*}[t]
\centering
\includegraphics[width=.9\textwidth] {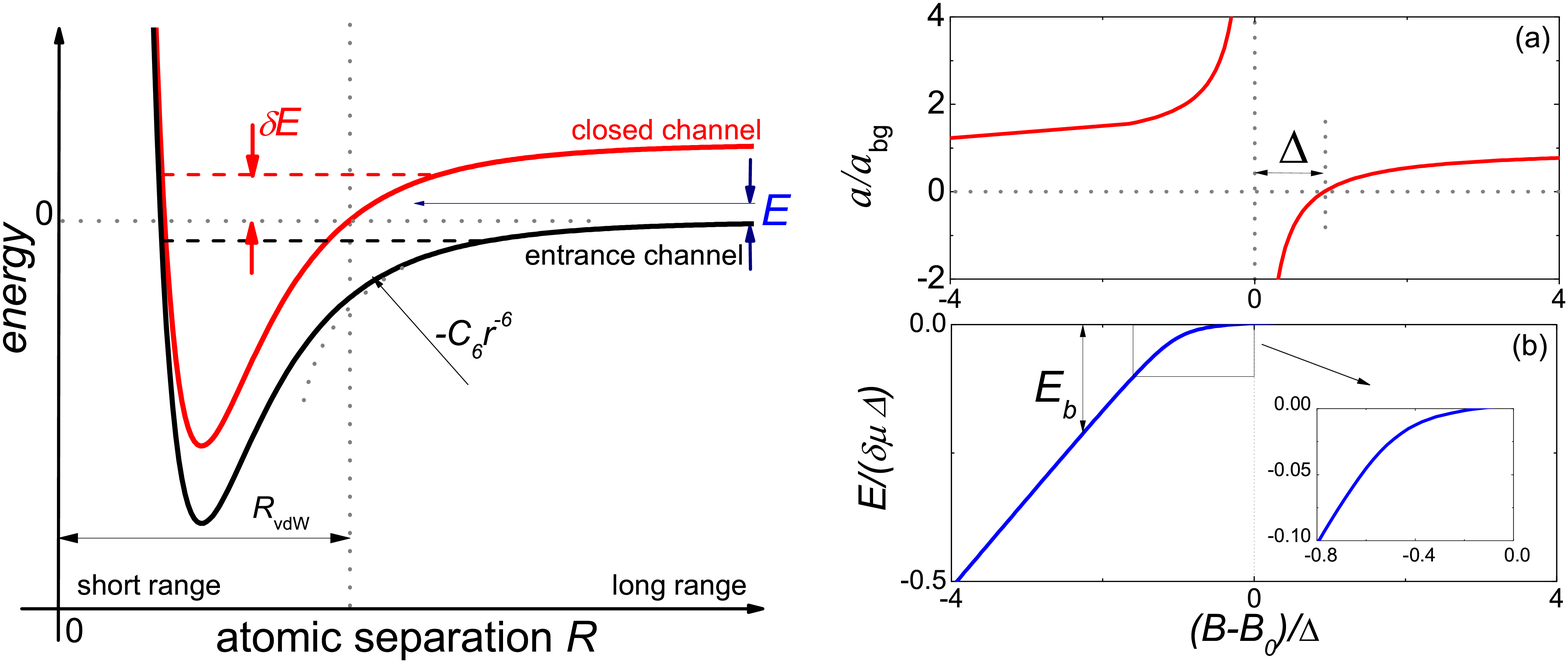}
\caption{Feshbach resonance phenomenon in ultracold collisions. Basic two channel model of a magnetically tunable Feshbach resonance is shown in the left-hand panel. Van der Waals interactions shape the long range potential and determine the boundary with the short-range physics. For ultracold collision with $E\rightarrow0$ the energy shift $\delta E$ between the free atom pair and the bound state (upper dashed line), supported by a different potential, can be tuned to resonance. The last bound state (lower dashed line) in the entrance channel is instead responsible of the background scattering length. Right-hand panel shows the divergent behavior of the scattering length at the pole the resonance (a) and binding energy (b) close to a Feshbach resonance. Note the quadratic behavior of the bound state in vicinity of the resonance pole.}
\label{Feshbach}
\end{figure*}

In the case of atoms in the electronic ground state, the long-range potential has a $-C_6/r^6$ tail governed by van der Waals interaction as depicted in Fig.\,\ref{Feshbach}. The constant $C_6$ is characteristic of the colliding atoms and its value can be numerically estimated from first principles \cite{Derevianko1999hpc}. The natural length scale associated with the van der Waals interaction is represented by the van der Waals length
\begin{equation}
R_{\rm{vdW}}=\frac{1}{2}\left (\frac{m C_6}{\hbar^2}\right)^{1/4},
\end{equation}
where $m$ is the mass of the atom and $\hbar$ is the reduced Planck constant. This parameter determines a natural boundary between short- and long-range physics in atomic collisions. 

\textit{Feshbach resonances} $-$ The scattering length between two particles (atoms, nucleons, molecules) is fixed by the details of the interactions and can have both positive and negative values. The exceptional power of ultracold atomic gases is the tunability of $a$ through magnetic Feshbach resonances \cite{Chin2010fri}. When a molecular state in a different interatomic potential (the \textit{closed channel}) has a different magnetic moment compared to the free atom pair (the \textit{entrance channel}), the relative energy can be tuned to zero; see Fig.\,\ref{Feshbach}. If the bound state then couples to the atomic threshold (e.g.\ by exchange coupling, spin-dipole or high-order spin-orbit), a divergence of $a$ is induced. The resonant behavior results in a large tunability of the scattering length from plus to minus infinity and in the case of a single isolated resonance, the dependence of the scattering length $a$ on the applied magnetic field strength $B$ can be parametrized in a convenient way as
\begin{equation}
a(B)=a_{\rm{bg}}\left (1-\frac{\Delta}{B-B_0}\right),
\label{fr}
\end{equation}
where $B_0$ and $\Delta$ are the pole and the width of the resonance, respectively, and $a_{\rm{bg}}$ is the background scattering length of the entrance channel (see Fig.\,\ref{Feshbach}).

\subsection{Two-body universality}
\label{uni}

The regime where the scattering length $a$ between two particles exceeds by far any other length scale of the two-body potential, $a\gg r_0, R_{\rm{vdW}}$, is known as \textit{universal regime}. The physics of the two-body system is then determined only by $a$ and becomes independent on the details of the short-range potential. At large positive values of the scattering length, a shallow two-body bound state is present below the atomic threshold. This state is universal and shows halo character as the wavefunction extends far into the forbidden region with only a small fraction of the probability density in the interaction region \cite{Jensen2004sar}. The dimer size is of the order of $a$ and the binding energy
\begin{equation}
E_b=\frac{\hbar^2}{ma^2}
\label{udimer}
\end{equation}
is quadratic in $1/a$. Near the center of the Feshbach resonance this corresponds to a quadratic dependence on the magnetic detuning $|B-B_0|$, as clearly visible in the inset of Fig.\,\ref{Feshbach}(b).

Although Eq.\,(\ref{fr}) holds for any isolated resonance, there is an important distinction between Feshbach resonances, which is based on the fraction of the width $\Delta$ over which universal behavior is present. This is commonly quantified by using the strength parameter $s_{\rm{res}}$, defined as \cite{Chin2010fri}
\begin{equation}
s_{\rm{res}}=\frac{a_{\rm{bg}}\bar{a}m}{\hbar^2}(\delta \mu\,\Delta),
\label{sreseq}
\end{equation}
where $\bar{a}=0.956R_{\rm{vdW}}$ is defined as the mean scattering length of the van der Waals interaction \cite{Gribakin1993csl} (for cesium $\bar{a} = 95.6\,a_0$) and $\delta \mu$ is the difference in magnetic moment between the closed and entrance channel. Large values ($s_{\rm{res}}\gg 1$) characterize \textit{open-channel dominated} resonances, where universal behavior is present for a large fraction of the resonance width. In this case, the scattering problem reduces to an effective single-channel problem with the spin character of the entrance-channel potential. Small values ($s_{\rm{res}}\ll 1$) define instead \textit{closed-channel dominated} resonances.

A useful quantity is the Feshbach resonance parameter $R^*\equiv \bar{a} / s_{\rm{res}}$ \cite{Petrov2004tbp}, which describes the length scale over which the Feshbach coupling influences the effective range $r_0$. On resonance ($a \rightarrow \pm \infty $) these two quantities are simply related by $r_0 = -2 R^*$. For small dimer binding energies the quadratic dependence on $1/a$ can be refined by introducing finite range corrections associated to $\bar{a}$ and $R^*$ \cite{Chin2010fri},
\begin{equation}
E_b=\frac{\hbar^2}{m (a-\bar{a}+R^*)^2}.
\label{hdimer}
\end{equation}
In the limiting case of $a\rightarrow\infty$, the energy relation for universal halo dimers is restored. The two possible terms of correction will be useful for the discussion on Efimov physics in Sec.\,\ref{sec:8}.
	
\subsection{Feshbach resonances in cesium}
\label{frcesium}

Ultracold cesium gases feature unique scattering properties, which have made them a prime candidate to observe Efimov physics. The atomic state of interest is the lowest internal state $|F\, = \,3, m_F\, = \,3 \rangle$, where $F$ and $m_F$ denote the quantum numbers for the hyperfine and magnetic state. Atoms in this state are immune against two-body inelastic decay, as there are no energetically open decay channels. The scattering properties of ultracold Cs have been subject to intense investigations for more than a decade. Feshbach spectroscopy \cite{Chin2010fri}, which is based on detecting resonances and investigating the underlying near-threshold molecular structure, has been intensively pursued first at Stanford University \cite{Chin2000hrf, Chin2004pfs} and then by our group in Innsbruck \cite{Mark2007sou, Berninger2011hmf}. Together with elaborate models for the near-threshold molecular structure \cite{Leo2000cpo, Chin2004pfs, Berninger2011hmf} a high level of understanding and accurate knowledge of the dependence of the scattering length on the magnetic field has been reached.

Two main factors make the Cs scattering properties special and largely different from many other species. First, the open channel features an extremely large scattering length, associated with the existence of a weakly-bound halo state. The corresponding background scattering length $a_{\rm bg} \approx +2200\,a_0$ exceeds the van der Waals radius $R_{\rm vdW}$ by more than a factor of 20, which according to Eq.~(\ref{sreseq}) strongly supports the occurrence of resonances with a large strength parameter $s_{\rm res} \gg 1$ and thus with a large universal range. Moreover, higher-order relativistic Feshbach coupling effects are particularly strong for this heavy species, enhancing the coupling to molecular states with higher partial wave character. Therefore, the Feshbach resonance structure in Cs is particularly rich and a zoo of extremely broad, intermediate and very narrow Feshbach resonances is found.

It is important to note that essentially all previous publications on Efimov physics in Cs published before 2011 relied on an $a(B)$ conversion that was based on the early Stanford experimental data and a theoretical model by NIST researchers \cite{Chin2000hrf, Chin2004pfs}. This model is quite accurate for low magnetic fields up to about 100\,G. Later our group largely extended the magnetic field range of Feshbach spectroscopy, and we performed additional measurements pinpointing zero crossings of the scattering length \cite{Zenesini2011CBE}. In collaboration with theorists from NIST (P. S. Julienne) and the Univ.\ or Durham (J. M. Hutson) this has resulted in an extended and an improved $a(B)$ conversion \cite{Berninger2011hmf}. We use the new $a(B)$ conversion throughout this article. For features in the low-field region, this results in small deviations for the specified scattering lengths in comparison to the originally published values.

\begin{figure*}[t]
\centering
\includegraphics[width=.85\textwidth] {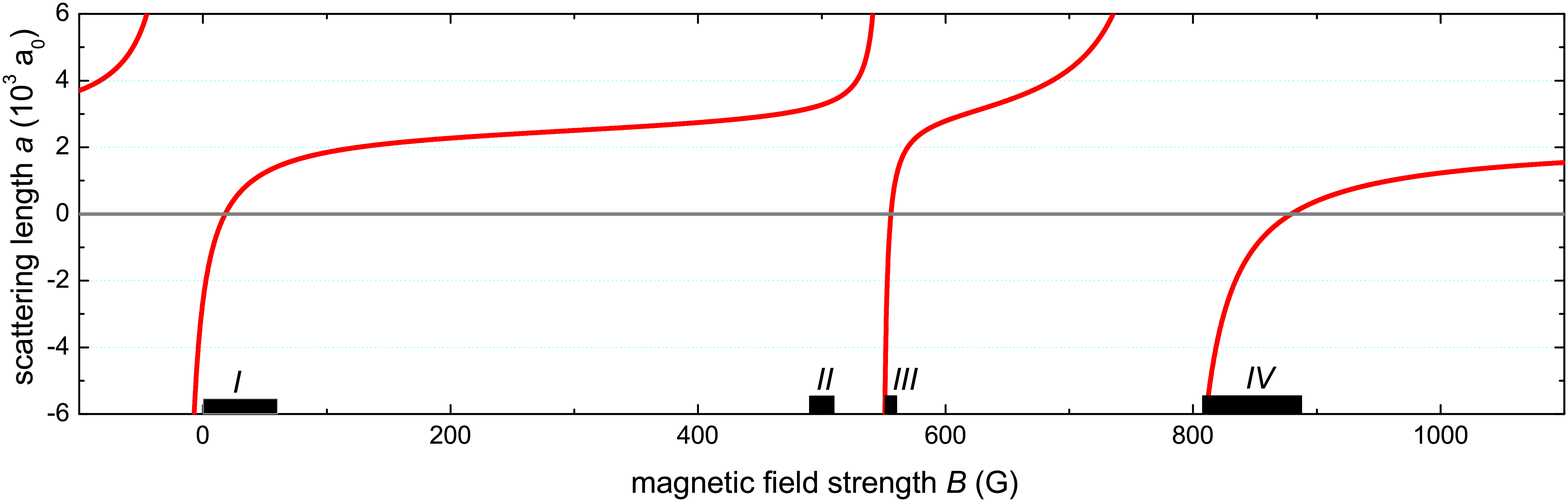}
\includegraphics[width=.85\textwidth] {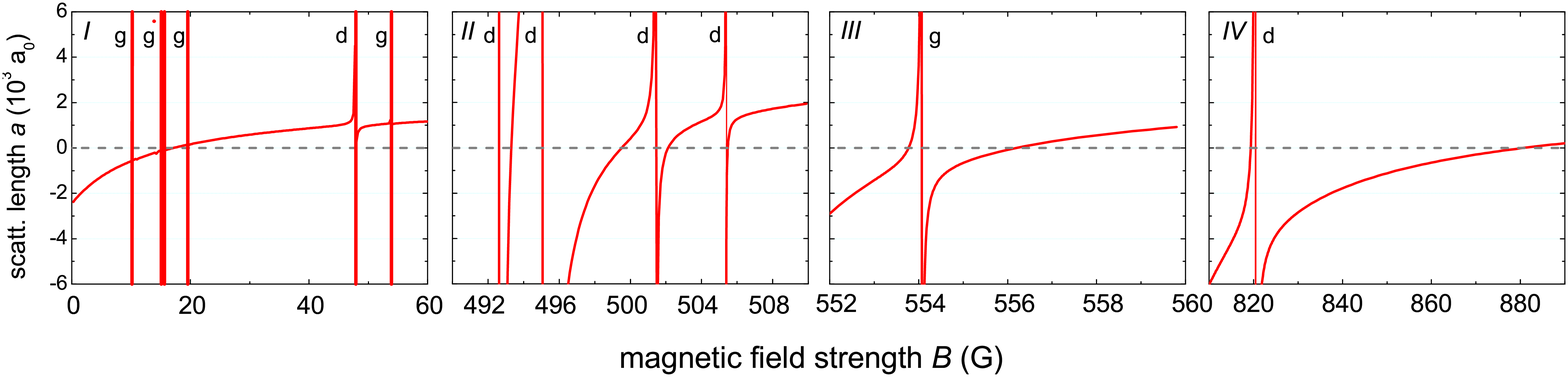}
\caption{Calculated magnetic field $B$ dependence of the two-body $s$-wave scattering length $a$ in the absolute ground state $|F\, = \,3, m_F\, = \,3 \rangle$ in $^{133}$Cs. (main panel) Overview of the $s$-wave scattering length in the accessible experimental region. Only contributions from molecular states with zero angular momentum are considered. (\textit{I-IV}) Zoom in of the relevant magnetic field regions including $s$-, $d$- and $g$-wave bound states. (\textit{I}) Feshbach resonance at $-12.3\,\rm{G}$ gives rise to a strong variation of $a(B)$ in the low magnetic field region. Several high order resonances sit both on positive and negative values of scattering length. (\textit{II}) Four $d$-wave resonance deeply shape the region around $500\,\rm{G}$, the broadest ones is centered at about $495\,\rm{G}$. (\textit{III}) The two overlapping $s$ and $g$ resonances near $550\,\rm{G}$. (\textit{IV}) The region between $800\,\rm{G}$ and $900\,\rm{G}$ is dominated by the broad $s$-wave resonance centered at $787.16\,\rm{G}$ and by the broad $d$-wave resonance at $820.37\,\rm{G}$. Letters in the different panels refer to the orbital angular momenta of the narrow resonances.}
\label{Scatt}
\end{figure*}

On a coarse-grain level the $s$-wave scattering properties are dominated by three broad $s$-wave Feshbach resonances with poles near $-12\,\rm{G}$, $549\,\rm{G}$ and $787\,\rm{G}$ \cite{Berninger2011uot, Chin2010fri, Chin2004pfs, Lee2007ete}.
The three resonances are generated by three molecular potentials with different spin properties. The fact that they are open-channel dominated ($s_{\rm{res}}$ exceeding one hundred) shows that they very well fulfill the conditions for universal behavior. Let us in the following discuss the different magnetic field regions where Efimov features have been observed in some more detail.

The low magnetic field region (region \textit{I} in Fig.\,\ref{Scatt}) has been subject of many investigations \cite{Chin2001hpf, Mark2007sou, Gustavsson2008coi}. The pole of the dominating Feshbach resonance is at $-12.3\,\rm{G}$ and the zero crossing has been determined with high accuracy to $17.130(2)\,\rm{G}$ by Bloch oscillation measurements \cite{Gustavsson2008coi,Gustavsson2008PhD}. The accessible range of scattering length is approximately between $- 2500\,a_0$ ($0\,\rm{G}$) and $+1500\,a_0$ ($100\,\rm{G}$) and this large tunability has been the key point for the first observation of Efimov features in 2006 \cite{Kraemer2006efe}.

The region of magnetic field around $500\,\rm{G}$ (region \textit{II} in Fig.\,\ref{Scatt}) is characterized by a cluster of four $d$-wave resonances. The widest one, centered at $495\,\rm{G}$ has a width of $4.5\,\rm{G}$ and has an open-channel dominated character, while the remaining three resonances are closed-channel dominated.

The Feshbach scenario around $550\,\rm{G}$ (region \textit{III} in Fig.\,\ref{Scatt}) shows two overlapping resonances, a broad $s$-wave and a narrow $g$-wave resonance. The presence of the wide $s$-wave resonance induces a negative background scattering length ($-1000\,a_0$) for the narrow $g$-wave resonance. This relative large background scattering length and the strong second-order spin-orbit coupling make this resonance a rare example of a broad $g$-wave resonance. Among the resonances presented here, this is the only one that is not closed-channel dominated and it represents a peculiar case as it resides on the shoulder of an open-channel dominated one\footnote{In this region of magnetic field, several $i$-wave resonances (neither plotted nor included in coupled-channel calculations) appear at the atomic threshold. The most relevant ones are at $557.45(3)\,\rm{G}$ ($\Delta\approx30\,\rm{mG}$) and $565.48(3)\,\rm{G}$ ($\Delta\approx40\,\rm{mG}$).}.

In the magnetic field region between $700\,\rm{G}$ and $900\,\rm{G}$ (region \textit{IV} in Fig.\,\ref{Scatt}) the Feshbach scenario features an extreme case of an open-channel dominated $s$-wave resonance with a width of $94\,\rm{G}$
\footnote{This resonance is an extreme case and has a larger $s_{\rm{res}}$ than any other known resonance in Cs or any other system.}. The tunability of the scattering length is very large on both sides of the resonance and the universal regime extends over many Gauss. An important additional feature is a broad $d$-wave resonance near 820\,G, which occurs at a very large and negative background scattering length of about $-4200\,a_0$.

\section{The Efimov effect}
	\label{sec:4}

The Efimov effect \cite{Efimov1970ela} is a fascinating and counterintuitive phenomenon occurring in a resonantly interacting three-body system. For extensive review on Efimov physics and related universal phenomena the reader may be referred to Refs. \cite{Jensen2004sar, Braaten2006uif}. Here we present the basic elements of Efimov's scenario and we discuss its particular impact on experiments with ultracold gases.

\subsection{The Efimov scenario}
\label{subsec:EfiScen}

\begin{figure}[h]
\begin{center}
\includegraphics[width=0.75\textwidth]{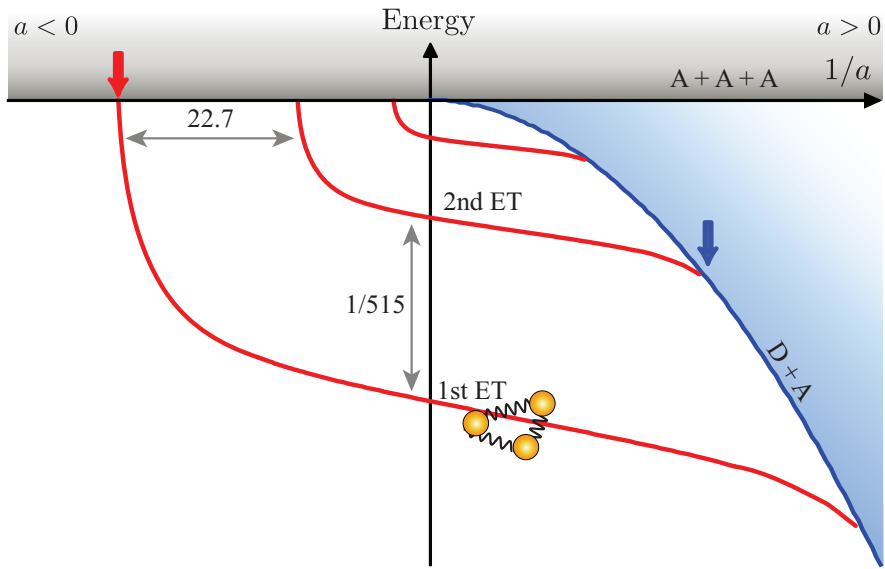}
\caption{{Efimov's scenario:} Infinite series
of weakly bound Efimov trimer states for resonant two-body
interaction. The binding energy is plotted as a function of the
inverse two-body scattering length $1/a$. The shaded regions  for $E>0$ and $E<0$ are the scattering continuum for three atoms  and
for an atom plus a dimer, respectively. The arrows mark the intersection of
the first Efimov trimer with the three-atom threshold and the atom-dimer threshold. To illustrate
the Efimov scenario just three Efimov states are depicted and the universal scaling factor was artificially reduced  from $22.7$ to $2$.} \label{Fig:Efimov_Scenario}
\end{center}
\end{figure}

Figure \ref{Fig:Efimov_Scenario} illustrates the Efimov scenario, consisting of a geometric series of  trimer states for large values of the scattering length $a$. The ladder of
three-body bound states is plotted versus the inverse scattering length $1/a$. Zero energy corresponds to the tri-atomic threshold, and for positive energy ($E>0$) the system consists of three free atoms with nonzero kinetic energy. Below the tri-atomic threshold ($E<0$), one can identify two different regions. For $a\!>\!0$, the pair-wise potential supports a universal weakly-bound dimer state with binding energy given by Eq.~(\ref{udimer}).
The corresponding threshold in the three-body picture is the atom-dimer threshold. The $a\!<\!0$ region is called the {\em Borromean region}, where counter-intuitively a series of three-body bound states can form although the two-body sub-systems are unbound. For $a\!<\!0$ the $n$-th trimer state crosses the tri-atomic threshold at $a_-^{(n)}$ and for $a\!>\!0$  the states merge with the atom-dimer threshold at $a_*^{(n)}$, where $n \ge 0$ is an integer number. With this convention, the Efimov state with largest binding energy is referred to as the first state and it is labelled with $n\!=\!0$. While this state crosses the tri-atomic threshold  at $a_-^{(0)}$, within {\em Efimov's window of universality} \footnote{Efimov's window of universality corresponds to the region $1/|a|\!\ll\!1/r_0$ and $E_b\!\ll\! \hbar^2/(m r_0^2)$.},
it may leave this window at the $a>0$ side, therefore losing its Efimov character. In this case, the state may not adiabatically connect to the atom-dimer threshold and the feature at $a_*^{(1)}$ may be the first atom-dimer resonance.

A remarkable property of the Efimov spectrum is the existence of a universal geometrical scaling law, governing the ladder of states. For three identical bosons, the geometrical factor is given by $e^{\pi/s_0}\!=\!22.7$, with $s_0\!=\!1.00624$. When increasing the length scale by a factor of 22.7, another Efimov state emerges at the tri-atomic threshold with a size 22.7 times larger. Furthermore, at $a\rightarrow \pm \infty$ the binding energies scale with $(22.7)^2\approx\!515$. The scaling factor fixes the relative energy between the trimers, while the absolute energy and the position of the first Efimov trimer is determined by an additional parameter, which is known as the {\em three-body parameter} (3BP). The 3BP enters into the theoretical descriptions as a cut-off to set a limit to the Efimov energy spectrum from below and consequently fixes the position $a_-^{(0)}$ of the first Efimov trimer state at the tri-atomic threshold; see Sec.\,\ref{sec:8}.

\subsection{Impact on ultracold gases}
\label{subsec:EfiLoss}

The success of ultracold gases in studying Efimov physics relies onto two main points. First, the high degree of tunability of the scattering length via magnetic Feshbach resonances offers unique possibilities to manipulate the interparticle interaction essentially ``on demand'' and thus to reach the universal regime in a well-controlled way.
Second, Efimov physics manifests itself in scattering properties in ultracold gases \cite{Nielsen1999ler,Esry1999rot,Braaten2006uif}. For instance, each time an Efimov trimer couples to a three-atom or to an atom-dimer threshold, the particle loss dramatically increases, and the corresponding scattering rate coefficients provide  well-suited observables to detect Efimov physics in experiments.

For atoms in the lowest internal state, i.e. the lowest Zeeman and hyperfine sublevel of the electronic ground state, inelastic two-body collisions are energetically suppressed and the dominant scattering and loss mechanisms come from three-body recombination processes. Here two atoms can bind into a deeply-bound state while the third atom ensures energy conservation. The  binding energy is converted into kinetic energy and the process generally results in losses of both the dimer and the third atom from the trapping potential that confines the ultracold sample.
As we will discuss in Sec.\,\ref{sec:9}, under certain circumstances also four-body processes can contribute significantly to the overall atom losses.

\begin{table}[t]
\caption{Universal connections between the positions of the tri-atomic Efimov resonance $a^{(n)}_{-}$, the atom-dimer Efimov resonance $a^{(n)}_{*}$, and the recombination-rate minimum $a^{(n)}_{+}$ \cite{Braaten2006uif}. The Efimov states are counted from $n\!=\!0$, meaning that the Efimov state with the largest binding energy crosses the tri-atomic threshold at $a^{(0)}_{-}$. Here, the symbol $a^{(n)}_{+}$ indicates the position of the $n$-th minimum in recombination in contrast to the notation previously used in Ref.\,\cite{Kraemer2006efe}.}
\centering
\label{tab:1}
\begin{tabular}{l l |c|c|c|}
\cline{3-5}
&&$a^{(n)}_{-}$ &$a^{(n)}_{+}$& $a^{(n)}_{*}$ \\ \hline
Position of the $(n+1)$-th tri-atomic Efimov resonance & \multicolumn{1}{||c|} {$a^{(n+1)}_{-}$} & $+22.7$ & $-22.7 \times 4.9$ & $-22.7/1.06$\\ \hline
Position of the $(n+1)$-th recombination minimum & \multicolumn{1}{||c|} {$a^{(n+1)}_{+}$} & $-22.7/4.9$ & $+22.7$ &$+4.4\times 22.7$\\ \hline
Position of the $(n+1)$-th atom-dimer Efimov resonance &\multicolumn{1}{||c|} {$a^{(n+1)}_{*}$} & $-1.06$ & +$22.7/4.4$ & $+22.7$\\ \hline
 \end{tabular}
 \end{table}

In an atomic sample, the Efimov effect impacts the three-body recombination in two opposite ways. For negative $a$, Efimov states induce pronounced {\em maxima} associated with a resonant increase of the recombination rate coefficient $L_3$ at the crossing position $a^{(n)}_-$, while for positive $a$ the recombination rate shows {\em minima}\footnote{Note that in Ref. \cite{Kraemer2006efe} $a^{(n)}_+$ denoted the maximum in $L_3$ at positive $a$. Here we use $a^{(n)}_+$ to indicate the minimum in $L_3$. } at $a^{(n)}_+$. The two positions are connected with each other according to the universal relations presented in Table \ref{tab:1} \cite{Nielsen1999ler,Esry1999rot}.
The recombination-rate maxima at $a<0$  result from the strong coupling  between three free atoms and the trimer state and the opening up of additional decay channels \cite{Esry1999rot}. These loss resonances are known as {\em tri-atomic Efimov resonances}. The recombination-rate minima at $a>0$ are the result of a destructive interference between two distinct decay pathways in the outgoing atom-dimer channel \cite{Nielsen1999ler,Esry1999rot,Greene2010uif}.

For large positive and negative values of $a$, the recombination rate coefficient $L_3$ takes the  simple form \cite{Weber2003tbr}
\begin{equation}\label{equ:lossParam}
L_3\!=\!3 C(a) \frac{\hbar a^4}{m} \,,
\end{equation}
which separates a general $a^4$ scaling \cite{Fedichev1996tbr,Nielsen1999ler} from a dimensionless log-periodic function $C(a)$ \cite{Braaten2006uif}, which contains the Efimov physics. A convenient quantity is the so-called three-body recombination length
\begin{equation}\label{equ:pho}
\rho_3=\sqrt[4]{\frac{2m}{\sqrt{3} \hbar}L_3},
\end{equation}
which converts the $a^4$-scaling of $L_3$ to a simple linear background dependence, $\rho_3\!=\!1.36\sqrt[4]{C(a)}|a|$ \cite{Esry1999rot}.

Effective field theory  provides analytic expressions for $C(a)$  for both negative and positive $a$  \cite{Braaten2006uif}
\begin{subnumcases}{C(a)=}
\label{equ:Cneg} 4590\,\frac{\sinh(2\eta_-)}{\sin^2[s_0\ln(a /a_-)]+\sinh^2\eta_-} & if $a<0$,\\
 \nonumber\\
67.1\,e^{-2\eta_+}(\sin^2[s_0\ln(a /a_+)]+\sinh^2\eta_+)+16.8\,(1-e^{-4\eta_+})& if $a>0$. \label{equ:Cpos}
\end{subnumcases}

The decay parameters $\eta_+$ and $\eta_-$ are related to the lifetime of the Efimov trimer. The trimer state can spontaneously decay into a deeply bound dimer state plus an atom. In a real atomic system, the two-body molecular spectrum is very rich and the inclusion of each two-body target state would make the problem very complex. The decay parameter absorbs the cumulative effect of the many decay channels. The most simple treatment is to consider that the decay parameter is constant for any value of the scattering length, i.\,e.\, that it assumes  the same value at both positive and negative scattering length ($\eta_-=\eta_+$). At a next level of complexity, one can consider a different value for the decay parameter for positive and negative $a$, as shown in Eqs.\,(\ref{equ:Cneg}) and (\ref{equ:Cpos}), or  even a decay parameter that varies with the scattering length \cite{Wenz2009uti}.
In the positive $a$ region, the effect of the deep states and the additional weakly-bound dimer is accounted for by the two additive terms in Eq.\,(\ref{equ:Cpos}).

\begin{figure}[h]
\begin{center}
\includegraphics[width=0.85\textwidth]{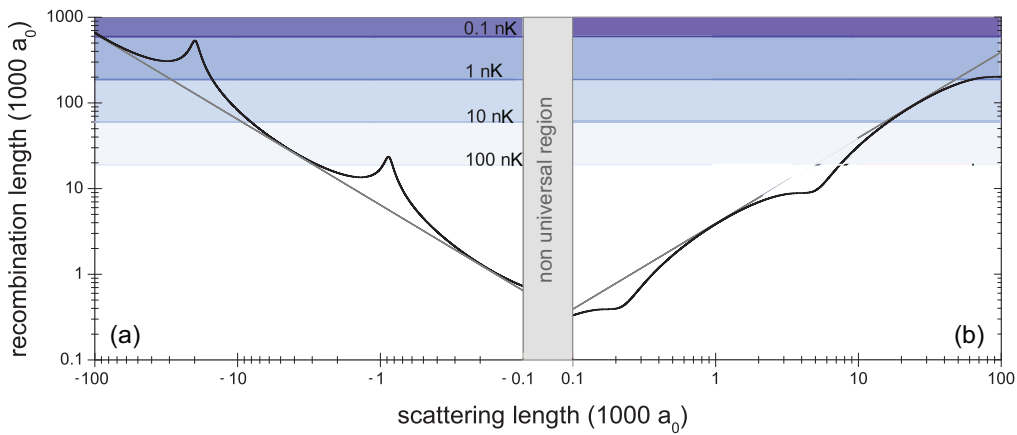}
\caption{Recombination length $\rho_3 $calculated within effective field theory for (a) negative and (b) positive values of the scattering length according to Eqs.\,(\ref{equ:Cneg}) and (\ref{equ:Cpos}), respectively. For the free parameters $a_{-(+)}$ and $\eta_{-(+)}$, we have chosen the values corresponding to the Cs Efimov features \cite{Kraemer2006efe}; see Table \ref{Efimov}.  The solid grey lines result from setting the $\sin^2$-terms to 1. The unitarity limited recombination length is also shown for 100, 10, 1 and 0.1~nK (from light to dark regions). } \label{Fig:Rho3}
\end{center}
\end{figure}

The function $C(a)$ has a log-periodic oscillatory behavior with $C(a) = C(22.7 a)$. It resonantly increases each time an Efimov trimer crosses the tri-atomic threshold at $a\!=\!a_-^{(n)}$ and decreases when the destructive interference occurs. Figures \,\ref{Fig:Rho3}(a) and (b) show the recombination length for negative and positive scattering length, respectively. For simplicity, we have omitted in the equations the index $(n)$, taking advantage of the log-periodicity of the $C(a)$ function.
The maximum and minimum position, $a_-$ and $a_+$, are universally connected accordingly to Table \ref{tab:1},  and just one of this two quantities has to be fixed to fully determine the three-body problem.  The resonant scattering length values and the $\eta$ parameter can not be  derived within the effective field theory, and need to be experimentally determined \cite{Wenz2009uti}.
As we will discuss in the next section, Eqs.(\ref{equ:pho}), (\ref{equ:Cneg}), and (\ref{equ:Cpos}) are commonly used to fit the experimentally measured values for the recombination rate, with all the four quantities $a_-$, $a_+$, $\eta_-$ and $\eta_+$ being left as free parameters\footnote{To analyze experiments the data for $a>0$ and $a<0$ are often fitted independently. The comparison of the results for $a_-$ and $a_+$ then provides a test for the universal relation.}.

For ultralow, but finite temperatures the three-body recombination length is unitarity limited  \cite{Dincao2004lou} to
\begin{equation}\label{equ:unilim}
 \rho_3^{\rm max}=5.2\frac{\hbar}{\sqrt{mk_BT}}.
\end{equation}
The unitarity limit imposes an upper value for the measurable recombination length. At $T\!=\!10$~nK, this limit corresponds to $\rho_3^{\rm max}\!=\!6\times 10^4\,a_0$ and at $T\!=\!100$~nK it is about a factor of three lower. As shown in Fig.\,\ref{Fig:Rho3}, the unitarity limit also sets the number of Efimov features that can be realistically observed in the experiments. While the first Efimov maximum is visible even for comparatively high temperatures ($T\simeq 100$~nK), the next maximum requires a temperature of about about a factor of 500 lower to be observable. Such a low temperature is hardly accessible in current experiments. This limitation  clarifies the reason why none of the experiments on Efimov physics was able to unambiguously  observe two consecutive maxima\footnote{Using an updated scattering length determination \cite{Gross2010nsi} the second maximum observed in Ref. \cite{Pollack2009uit}  may be interpreted as resulting from the pole of the Feshbach resonance and not from an Efimov state.}, although two neighboring minima have been successfully revealed \cite{Zaccanti2009ooa,Pollack2009uit}; see Sec.\,\ref{sec:10others}.

The above discussion has focused on the three-body behavior at the tri-atomic threshold.  Similar arguments apply to the atom-dimer threshold ($a>0$). When an atom-dimer system resonantly couples to an Efimov trimer the scattering rate resonantly increases. Here the dominant loss mechanism is collisional relaxation, also known as vibrational quenching. During an atom-dimer collision, the dimer can relax into a more deeply bound two-body  state and the binding energy is converted into kinetic energy. This two-body process is described by the rate equation $\dot{n}_{\rm D}\!=\!\dot{n}_{\rm A}\!=\!-\beta n_{\rm D} n_{\rm A}$, where $n_{\rm D (A)}$ is the dimer (atom) density and $\beta$ is the relaxation-rate coefficient.
At the atom-dimer Efimov resonance position $a\!=\!a_*$ the relaxation rate exhibits a resonant increase.
Effective field theory provides a universal formula for the relaxation-rate coefficient $\beta$ at zero temperature \cite{Braaten2004edr, DIncao2005sls, Braaten2007rdr}
\begin{equation}\label{eqn:ADBeta}
\beta= C_{\rm AD}(a) \frac{\hbar a}{m},
\end{equation}
where $C_{\rm AD}(a)$ incorporates the typical log-periodic oscillator behavior of the Efimov effect and reads as
\begin{equation}\label{eqn:ADC}
C_{\rm AD}(a)= \frac{20.3\sinh(2\eta_*)}{\sin^2\left[s_0\ln(a/a_*)\right]+\sinh^2\eta_*}.
\end{equation}
The parameter $\eta_*$ is again related to the trimer lifetime. Within effective field theory $\eta_*\!=\!\eta_-\!=\!\eta_+$, while for convenience all these three parameters are treated as independent when fitting the theoretical expression to experimental results.

\section{Efimov resonances in ultracold quantum gases of Cs atoms}
 	\label{sec:5main}

In summer 2005, an Efimov trimer state was observed in experiments performed by our group in Innsbruck \cite{Kraemer2006efe}. These experiments, conducted on optically trapped ultracold gases of Cs atoms, provided signatures of both a tri-atomic Efimov resonance and a recombination minimum. Later experiments showed the appearance of an atom-dimer Efimov resonance \cite{Knoop2009ooa}. Figure \ref{fig:efilow} summarizes the main observations of Efimov resonances in Cs for both the $a<0$ and the $a>0$ side.

All these experiments were based on the magnetically tunable interaction properties of Cs atoms in a magnetic field range between $0$ and $150\,\rm{G}$, henceforth referred to  as the \emph{low-field region} (region \textit{I} in Fig.\,\ref{Scatt}). The accessible magnetic field range was technically limited by our previous magnetic coil setup, restricting the tunability of the $s$-wave scattering length $a$ to a range between $-2500\,a_0$ and $1600\,a_0$. In 2010, we performed a major upgrade of our setup, which now allows us to produce magnetic fields of up to 1.4~kG. With this new setup, we have studied Efimov physics in the $550$ region and in the  $800\,\rm{G}$ region \cite{Berninger2011uot}, henceforth referred to as the \emph{high-field region}, where two broad $s$-wave Feshbach resonances allow for a wide tunability of the scattering length \cite{Berninger2011hmf}.

In this Section, we first present our observations on the  tri-atomic Efimov resonances in both the low-field and the high-field region (Sec.\,\ref{sec:5}) and on the atom-dimer Efimov resonance observed in the low-field region (Sec.\,\ref{sec:6}). We then describe further Efimov resonances observed on other Feshbach resonances in Cs  (Sec.\,\ref{sec:gEfi}), and we discuss the behavior of the three-body parameter in our observations (Sec.\,\ref{sec:8}).
	
\subsection{Triatomic Efimov resonances}
 \label{sec:5}

To reveal the Efimov resonances at the tri-atomic threshold, we prepare an optically trapped thermal sample of up to $5 \times 10^4$ Cs atoms at temperatures ranging from 10 to 250~nK. Our experimental procedure is based on an all-optical cooling approach as presented in Refs.\,\cite{Weber2003bec,Kraemer2004opo,Rychtarik2004tdb}.
The atoms are prepared in their lowest internal spin state ($F=3, m_F=3$), where three-body recombination collisions are the dominant loss mechanism.
We measure the three-body loss rate $L_3$  by recording the time evolution of the atom number $N$ and the atom temperature $T$ for different magnetic field values in the region of interest \cite{Weber2003tbr, Kraemer2006efe}.
In the following, we discuss separately the recombination results obtained in the low- and high-field region.

\begin{figure*}
\centering
\includegraphics[width=0.85\textwidth]{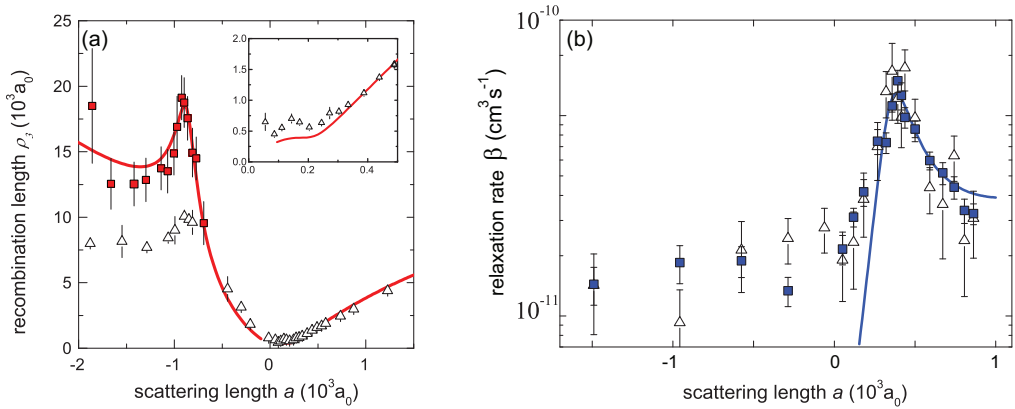}
\caption{ Efimov resonances observed in ultracold cesium at low magnetic fields. {(a)} Observation of a triatomic Efimov resonance in measurements of
three-body recombination. The recombination length $\rho_3$ is plotted as a function of the scattering length $a$.
The squares and the empty triangles show the experimental data for initial temperatures around 10\,nK, and 200\,nK, respectively \cite{Kraemer2006efe}. The solid curve
represents the analytic model from effective field theory
\cite{Braaten2006uif}, where the scattering-length to magnetic-field conversion is from \cite{Berninger2011uot}. The inset shows an expanded view for small
positive scattering lengths with a minimum near $210\,a_0$. The displayed error bars refer to
statistical uncertainties only. {(b)} Two-body loss rate coefficient $\beta$ measured for inelastic atom-dimer collisions \cite{Knoop2009ooa} at two different temperatures, 40\,nK (open triangles) and 170\,nK (filled squares). Here a prominent atom-dimer Efimov resonance shows up for $a>0$. The solid lines represent fits based on universal effective field theory.}
\label{fig:efilow}
\end{figure*}
\vskip 0.5 cm
\noindent
\emph{\bf Low-field region}
\vskip 0.2 cm
\noindent
We first focus our attention on the low magnetic-field region and discuss the results of Ref.\,\cite{Kraemer2006efe}\footnote{The data have been refitted according to the new $a(B)$ conversion.}. As shown in Fig.\,\ref{Scatt} (region \emph{I}), the smooth change in  the scattering length is caused by an $s$-wave Feshbach resonance located at about $-12$~G. The scattering length passes from large negative to large positive values through a zero crossing, located at about 17~G,  rather than through a pole.

The data are plotted in Fig.\,\ref{fig:efilow}(a) in terms of the recombination length $\rho_3$; see Eq.\,(\ref{equ:pho}).
In the negative $a$ region we observed a prominent triatomic Efimov resonance as a dramatic increase of the three-body recombination rate. As expected from effective field theory, the recombination length $\rho_3$ shows a resonant increase on top of a linear dependence on $a$.
Our data follow remarkably well the functional form of $L_3$ derived with the effective field theory at zero temperature. From the fit of Eq.\,(\ref{equ:pho}) to our 10~nK experimental data we extract the resonance peak position to be $a_-^{(0)}=-872(22)\,a_0$, corresponding to about 7.6~G, and the decay parameter $\eta_-=0.10(3)$.

All the results discussed so far are valid in the zero-energy collision limit. This approximation is well justified for our lowest temperature data ($T=10$~nK) where the largest measured recombination-rate coefficient is well below the upper bound imposed by the unitarity limit.
At higher temperatures, the unitary limit affects the observability of the Efimov resonance; see Eq.\,(\ref{equ:unilim}).
This effect is clearly visible in Fig.\,\ref{fig:efilow} where both 10 and 250~nK data are shown.
For 250\,nK, the resonance feature is still visible (open triangles) but with about half the contrast of the low temperature data. In further experiments at higher temperatures (data not shown) we observed the resonance to disappear above $\sim$500\,nK.
With increasing temperature we also observed a linear shift of the resonance position to lower $|a|$ values  with a slope of $0.2\, a_0/{\rm nK}$ \cite{Naegerl2006eef}, which confirmed the picture of an Efimov state crossing the tri-atomic threshold. Above threshold the trimer state is expected to become metastable and the zero-energy Efimov resonance evolves into a triatomic continuum resonance \cite{Bringas2004tcr,Yamashita2007tbr}.

In Ref.\,\cite{Kraemer2006efe}, we also investigated the positive $a$ region, which is connected to the negative one through a zero crossing. Here we observed a recombination minimum at $a_{+}^{(0)}\approx 210\,a_0$; see inset to Fig.~\ref{fig:efilow}.  We fitted Eq.\,(\ref{equ:pho}) to our $a>0$ recombination data and we observed a good agreement with the results from effective field theory. Although the measured $a_-^{(0)}/a_+^{(0)}$ ratio is in good agreement with the predicted universal value (see Table \ref{tab:1}), the interpretation of the minimum in terms of universal theory is a delicate issue as the minimum appears at a value of $a$ that is only about two times larger than $R_{\rm vdW}$. In the regime of comparatively small scattering lengths, non-universal corrections might play a sizable role, affecting the universal ratio values.
From the theoretical side,  the relation between features occurring at opposite sides of a zero crossing of the scattering length have risen controversial interpretations. In Ref.\,\cite{Dincao2009tsr}, the authors rose serious doubts about the possibility to universally relate the observed minimum to the tri-atomic Efimov resonance. In Refs.\,\cite{Lee2007ete,Platter2008sfa,Massignan2008esn,Jonalasinio2010tru} the specific case of Cs at low magnetic field was considered. The calculations are all consistent with the appearance of a recombination-rate minimum at about $200\,a_0$, as observed in our experiments.

The open issues in the interpretation of our low-field results has been a main motivation to extend our measurements to the high-field region.

\begin{figure}[b]
\centering
\includegraphics[width=1\textwidth] {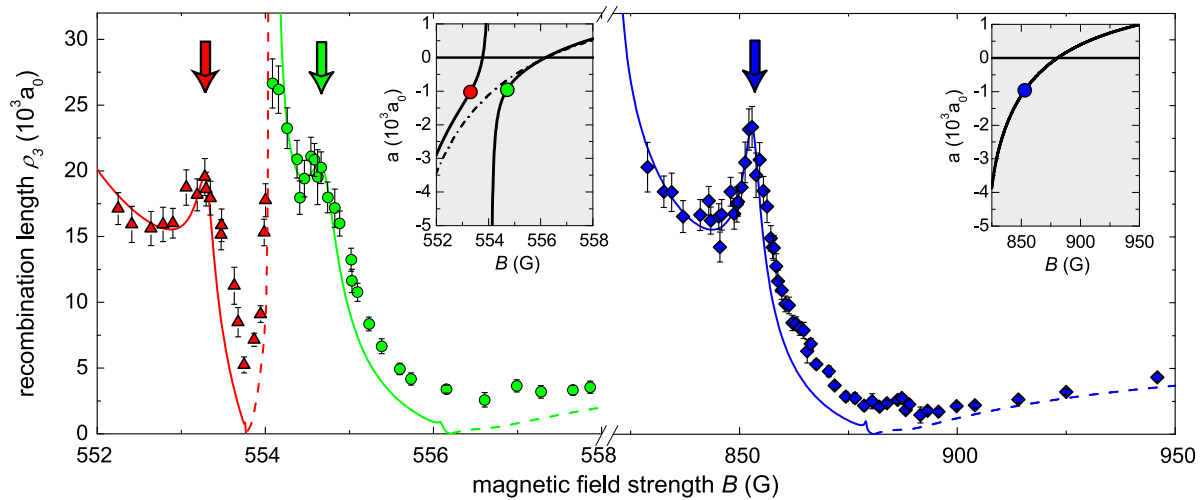}
\caption{Efimov resonances at high field. The measured recombination length $\rho_3$ is shown for three different regions (triangles, $552\,\mathrm{G} < B < 554\,\mathrm{G}$; bullets, $554\,\mathrm{G} < B < 558\,\mathrm{G}$; lozenges, $830\,\mathrm{G} < B < 950\,\mathrm{G}$), which are separated by the poles of different Feshbach resonances. The error bars indicate the statistical uncertainties. For all three regions, the solid lines represent independent fits to the data at negative $a$.
The dashed lines show the predictions of effective field theory for $a>0$ \cite{Braaten2006uif}, using the parameters obtained in the corresponding region for $a<0$. The insets show the relevant behavior of the scattering length (solid line, full calculation; dash-dot line, $s$-wave states only). The arrows in the main figure and the corresponding dots in the insets refer to the triatomic Efimov resonances.}
\label{swave550}
\end{figure}

\begin{figure}[t]
\centering
\includegraphics[width=.85\textwidth] {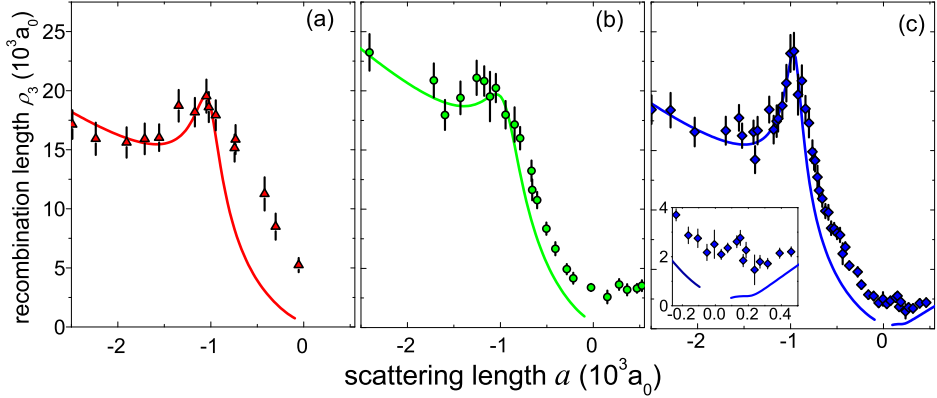}
\caption{Comparison of triatomic Efimov resonances in the high-field region of Cs. The data are the same as in Fig.\,\ref{swave550}, but here a $\rho_3(a)$ representation facilitates a direct comparison. Recombination length $\rho_3$ versus scattering length near the $553.3(4)\,\rm{G}$ (a) and $554.71(6)\,\rm{G}$ (b) three-body recombination maxima. (c) Recombination length $\rho_3$ versus scattering length for the $853.07\,\rm{G}$ three-body recombination maximum. Inset in (c) Zoom on the recombination minimum at $270(20)\,a_0$.}
\label{summary}
\end{figure}

\vskip 0.5 cm
\noindent
\emph{\bf High-field region}
\vskip 0.2 cm
\noindent
Here we summarize our recent observations on Efimov resonances in the high-field region, as presented in Ref. \cite{Berninger2011uot}. The measurement procedure was basically the same as for the low-field case. In Fig.\,\ref{swave550} we plot the measured recombination lengths $\rho_3$ versus magnetic field strength $B$, while in Fig.\,\ref{summary} the same data are plotted for all observed Efimov resonances as a function of the scattering length, according to the $a(B)$ conversion.

In the 800\,\rm{G} (region \textit{IV} in Fig.\,\ref{Scatt}), the broad resonance represents  an extreme case of an open-channel dominated resonance ($s_{\rm{res}}\approx1470$) and it has strong similarities to the low-field $s$-wave resonance. In the region of negative scattering length, measurements reveal a feature with a width of a few G. Its shape closely resembles the results of the low-field region and a fit to the data according to the Eq.\,(\ref{equ:Cneg}) gives a value of $a_- = -955(28)\,a_0$. This shows that an Efimov state appears at the atomic threshold at a value for $a$ very close to the low-field one. Furthermore, at even higher magnetic field, beyond the zero crossing of the scattering length, we found a loss minimum at $B = 893(1)$\,G, corresponding to $a_+ = 270(30)\,a_0$; see inset of Fig.\,\ref{summary}(c). These two features closely resemble the ones observed at low field and the ratio between the positions of the maximum and the minimum is consistent with universal theory. Future investigations will aim at the measurement of the second recombination minimum expected according to universal theory around $+4300\,a_0$.

\begin{table}[b]
\begin{center}
\caption{Overview of observed Efimov features in ultracold Cs. For the corresponding Feshbach resonances (see also Fig.\ \ref{Scatt}) we specify the corresponding partial-wave character of the molecular state (second column) and we give the resonance strength parameter (sixth column). The third and fourth column give the positions of the triatomic Efimov resonances in terms of scattering length and magnetic field, respectively. The fifth column specifies the decay parameter $\eta_-$. The last column indicates the positions of the observed recombination minima. The values for $a_-$ and $\eta_-$ were obtained by universal fits, whereas the given values for $a_+$ were directly from the positions of the corresponding loss minima. ($^{\$}$Note that the marked values may suffer from systematic errors larger than the specified uncertainty; see discussion in Sec.~\ref{sec:gEfi}.)}
\begin{tabularx}{14cm}{c|c||c|c|c|c|c}
~ Region ~ &
partial wave&~ $B(a_-) (\rm{G})$ ~&~ $a_- (a_0)$~&~ $\eta_-$~&~ $s_{\rm{res}}$ ~~&~ $a_+(a_0)$\\
\hline
~\textit{I}~ &~$s$~&~ 7.56(17) ~&~ -872(22) ~&~ 0.10(3)~&~ 560 ~&~{210(10)}~\\
\hline
~\textit{II}~ &~$d$~&~ $498.1(1)$~&~ $-$ ~&~ $-$~&~ $\approx100$~&~ $-$~ \\
\hline
~\multirow{2}{*}{\textit{III}}~&~$s$~&~ 554.71(6) ~&~ -957(80) ~&~ 0.19(2) ~&~ 170 ~&~ {250(40)}~\\
&~$g$~&~ 553.30(4) ~&~ -1029(58) ~&~ 0.12(1) ~&~ 0.9 ~&~ $-$~\\
\hline
~\multirow{2}{*}{\textit{IV}}~&~$d$~&~ 818.89(7)$^{\$}$ ~&~ -1400(150)$^{\$}$~&~ 0.18(3)$^{\$}$ ~&~ 12~&~ $-$~ \\
&~$s$~&~ 853.07(56)~&~ -955(28) ~&~ 0.08(1) ~&~ 1470~&~ {270(30)}~
\end{tabularx}
\end{center}
\label{Efimov}
\end{table}

Cesium features a second broad $s$-wave resonance between near $548\,\rm{G}$; see region \textit{III} in Fig.\,\ref{Scatt}. Here, a $g$-wave resonance centered at $554.4\,\rm{G}$ creates a more complicated scenario as it resides on the shoulder of the $s$-wave resonance at about $-1000\,a_0$. This value is very close to the position where one may expect an Efimov resonance to occur if only $s$-wave contributions were considered (left inset in Fig.\,\ref{swave550}). The presence of the $g$-wave resonance generates two regions of negative $a$ with a range of $300\,\rm{mG}$ of positive $a$ in between. As discussed in Sec.\,\ref{sec:3}, while the $s$-wave resonance is open-channel dominated ($s_{\rm{res}}\approx170$), the character of the $g$-wave resonance is an intermediate case ($s_{\rm{res}}=0.9$). Our $L_3$ measurements show a loss minimum at $553.7\,\rm{G}$ resulting from the zero crossing of the $g$-wave resonance, and additionally three loss peaks are clearly observed (see left-hand side of Fig.\,\ref{swave550}). The central peak is the $g$-wave resonance pole, while we interpret the other two features as triatomic Efimov resonances. The separate fit to the two loss features with the universal theory gives $a_-$= $-1029(58)\,a_0$ for the left feature and $-957(80)\,a_0$ for the right one. Even in this scenario of overlapping resonances, the values of $a_-$ are close to the ones measured at low fields and in the 800\,G region. The $L_3$ data on the lower side of the $g$-wave resonance pole may show experimental artifacts resulting from magnetic field ramping, and they may also be influenced by a possible magnetic field dependence of $\eta_-$ \cite{Wenz2009uti}. Future investigations could consider different ramping schemes of the magnetic field in order to separate the different contributions.
In additional sets of measurements (not shown) we also detected an Efimov recombination minimum around $556.9(2)\,\rm{G}$, corresponding to $a_+ = 250(40)\,a_0$, which again fits to the other observations and a universal connection of all features.

In Table\,\ref{Efimov} we summarize all our observations on Efimov-related features in atomic three-body recombination observed in ultracold Cs. All observations support a universal connection. This in particular means that the three-body parameter stays essentially constant as we will discuss later on in Sec.\,\ref{sec:8}.

\subsection{Atom-dimer resonances}
 	  \label{sec:6}

The Efimov scenario predicts that the trimer states merge with the atom-dimer threshold for $a>0$. This causes resonant increase of the relaxation rate in atom-dimer collisions  \cite{Braaten2004edr}. In Ref.\,\cite{Knoop2009ooa} we reported the first observation of an atom-dimer resonance. We produced a  mixture of about $3\times 10^4$ Cs atoms and $4\times 10^3$  weakly-bound Cs$_2$ halo dimers \cite{Mark2007sou,Ferlaino2009ufm,Chin2010fri}.
The halo dimer state used in the experiments is a universal $s$-wave bound-state with $E_b<E_{\rm vdW}\simeq h \times 2.7$ MHz\footnote{$E_{\rm vdW} = \hbar^2/(m R^2_{\rm vdW})$ is the van der Waals energy \cite{Chin2010fri}} for magnetic field values $B$ above 20~G. For $B<20$~G, the binding energy becomes too large and the state loses its universal character \cite{Mark2007sou}.

The relaxation loss rates were measured by probing the decay of the number of optically trapped dimers for varying magnetic field strengths. To investigate the role of the initial collisional energy, the measurements were performed both on a 40\,nK and a 170\,nK mixture. Figure \ref{fig:efilow}(b) shows the atom-dimer relaxation rate coefficients $\beta$ as a function of the scattering length $a$ for both temperature settings.
The  data clearly show a loss resonance that we interpret as an  atom-dimer Efimov resonance. By increasing the scattering length to large positive values the relaxation rate first undergoes a rapid increase of about an order of magnitude and then a smooth decrease toward its typical background value. We observe similar relaxation-rate values for both the low- and high-temperature data set.
From the fit of Eq.\,(\ref{eqn:ADC}) to the $170$-nK data, we determined the atom-dimer resonance position to $a_*^{(1)}\!=\!+367(13) a_0$, corresponding to about 25~G, and the decay parameter to $\eta_*=0.30(4)$. We found an overall good qualitative agreement with the predictions of effective-field theory at zero temperature. However, we always measured smaller relaxation rate coefficients than the value expected at $T=0$  \cite{Braaten2004edr}.
The observed behavior can partially be explained considering finite-temperature effects \cite{Knoop2009ooab}.
References \cite{Braaten2007rdr,Braaten2009erd,Helfrich2009rad} report on an extension of  the effective field theory for non-zero temperature. The authors predicted a comparatively rapid decrease of $\beta$ with increasing temperature. In Ref.\,\cite{Helfrich2009rad}, the absolute values of $\beta$  calculated for $T=170$~nK show a good agreement with our high-temperature data. However the theory predicted a temperature dependence that  could not be observed in the experiments. This could be due to systematic problems with our low-temperature data or to more fundamental reasons. Further investigations are needed to clarify this point.

Another important issue concerns the universal connection between the tri-atomic and the atom-dimer Efimov resonance.
As reported in Table \ref{tab:1}, the universal theory predicts a ratio  $a^{(n+1)}_{*}/|a^{(n)}_{-}|\approx 1.06$, while we found $a^{(1)}_{*}/|a^{(0)}_{-}|\approx 0.4$. The observed deviation may be partially connected with the fact that the condition for universality ($a\!\gg\!R_{\rm vdW}$) is not very well fulfilled since the  atom-dimer resonance occurs at a value of the scattering length that is only a factor 3.6 larger that $R_{\rm vdW}$.
Non-universal and finite-range contributions could in fact  modify the scaling ratio. It is worth to mention that the other two experiments on atom-dimer resonances, one with $^{39}$K \cite{Zaccanti2009ooa}  and the other one with $^7$Li \cite{Pollack2009uit}, reported
similar deviations from the universal ratio. All these results rise the question whether the deviation is purely accidental or whether it points to some fundamental issues that still need to be understood and explored. Future experiments on atom-dimer resonances in Cs at high magnetic fields could  help to shed some new light on this issue.

\subsection{Evidence for further Efimov resonances}
 	  \label{sec:gEfi}

In addition to the broad Feshbach resonances discussed before, Cs offers many more narrower resonances. One therefore may expect further Efimov features to be present, only that they are more difficult to observe and to interpret in terms of the scattering length. On the other hand, such features may be of great interest to clarify the situation for Feshbach resonances that are not in the open-channel dominated limit \cite{Petrov2004tbp}. Here, we present first results on Efimov features on two resonances that originate from $d$-wave molecular states.

\begin{figure}[t]
\centering
\includegraphics[width=.8\textwidth] {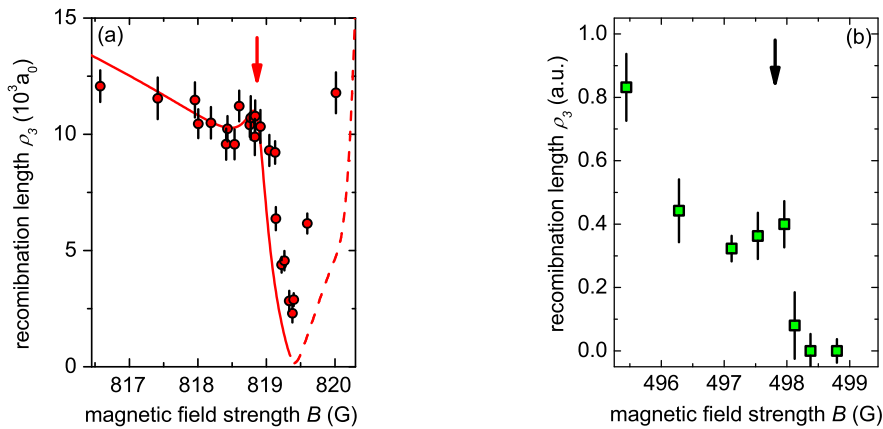}
\caption{Efimov features close to $d$-wave Feshbach resonances. The recombination length $\rho_3$ is plotted vs magnetic field strength. The error bars come from  statistical uncertainty on trap frequency, temperature and recombination decay rate. (a) Efimov resonance of the $820\,\rm{G}$ $d$-wave Feshbach resonance. The fitting parameters in the negative region of $a$ are $a_-=1400(150)\,a_0$ and $\eta_-=0.18(3)$, while the positive region the line considers $\eta_-=\eta_+$ and $a_-=a_+/4.9$, as Table\,\ref{tab:1}. (b) Data from three-body decay rate reveal an Efimov feature near the $495\,\rm{G}$ $d$-wave Feshbach resonance.  An accurate number density calibration for panel (b) is not available and therefore arbitrary units are used for the vertical axis.}
\label{dwave800}
\end{figure}

As shown in region in Fig.\,\ref{Scatt} the broad $s$-wave Feshbach resonance at $787\,\rm{G}$ presents an overlapping $d$-wave resonance, which is moderately entrance-channel dominated ($s_{\rm{res}}\approx 12$). As the resonance is found on the shoulder of the $s$-wave resonance at a large value of scattering length ($-4200(200)\,a_0$), the background scattering length is rapidly changing across the width of the $d$-wave resonance and this could modify the three-body recombination rate \cite{Esry2011pc}.
Below the $d$-wave zero-crossing at $819.37(3)\,\rm{G}$ we observe a peak structure not connected to other Feshbach resonances as shown in Fig.\,\ref{dwave800}(a). We assign it to an Efimov feature and by using universal theory, we obtain $a_-=-1400(150)\,a_0$ for this feature. Our $a(B)$ conversion is much less precise in this region and the given error bar does not include systematic uncertainties, which for this case may substantially exceed the statistical uncertainties. Possible systematic error sources may be systematic, model-dependent uncertainties in fitting the near-threshold molecular structure and experimental, magnetic-field ramping issues, which may somewhat distort the observed structure. Therefore this value for $a_-$ cannot be interpreted as significantly different from the ones reported before.

Further evidence for an Efimov state on a $d$-wave Feshbach resonance has been found at about $498\,\rm{G}$ in the negative $a$ region of the strongest $d$-wave resonance. This open-channel dominated resonance ($s_{\rm{res}}>100$) is influenced by other three nearby $d$-wave resonances that modify the background scattering length in that region of $B$ (region \textit{II} in Fig.\,\ref{Scatt}). A loss resonance is clearly visible in $L_3$ measurements plotted in Fig.\,\ref{dwave800}(b) and we assign this loss feature to an Efimov state. The poor statistics and systematic errors, as discussed for 820\,G feature,  do not allow a precise determination of $a_-$ and $\eta_-$. Nevertheless, both these additional observations highlight the ubiquitous nature of Efimov states in bosonic gases with large scattering length, and in future experiments their quantitative investigation may provide more insight into the particular role of the Feshbach resonances.

\subsection{Three-body parameter}
	   \label{sec:8}

Efimov's geometrical scaling law (Sec.~\ref{sec:4}) determines the relative energy between the trimer states. To fix their absolute positions, however, an additional parameter is needed. This parameter is commonly referred to as the \textit{three-body parameter} (3BP). The 3BP corresponds to a cut-off of high lying states in momentum space, which bounds the ladder of Efimov states from below. For zero-range models, this avoids the unphysical situation of bound states with an infinitely large binding energy (known as the Thomas effect \cite{Thomas1935tib}). The physical meaning of the 3BP in real systems and the relation to the short-range physics is still an open issue. Here we discuss possible interpretations of this quantity, and we discuss experimental findings shedding light on this issue.

The general idea is that the 3BP depends on all the details of the two- and three-body potentials not captured by universality. Short-range physics enters via two-body interactions, but also through genuine three-body terms, which are a non-additive contributions to the total interaction potential \cite{Bedaque1999rof}.  D'Incao and coauthors \cite{Dincao2009tsr} numerically demonstrated the possible effect of a three-body model interaction on the recombination rate. However, as shown in \cite{Soldan2003tbn}, a realistic three-body interaction modifies the system behavior at a length scale much smaller than $R_{\rm{vdW}}$\footnote{Typically genuine three-body interactions have a range of $10\,a_0$, which is one tenth of our $R_{\rm{vdW}}$.}. In the universal limit, a Feshbach resonance only modifies the part of the wave function that is far outside of the van der Waals range, which suggests a negligible contribution of the three-body forces to the 3BP. In fact, some recent theoretical approaches \cite{Lee2007ete, Jonalasinio2010tru, Naidon2011tee} obtained the positions of Efimov features approximately right without invoking any three-body forces. At present, the possible impact of three-body interactions in real experiments with ultracold atoms remains an issue that is not fully resolved.

If we assume that the role of three-body forces is negligible, then the problem can be fully described in terms of pair-wise two-body interactions. Here the main question is what sets the length scale associated with the 3BP, as represented by the scattering length $a^{(0)}_-$ where the first Efimov resonance occurs. For closed-channel dominated resonances, $R^* \gg \bar{a}$ provides an additional length scale \cite{Petrov2004tbp, Hammer2007erc, Massignan2008esn, Pricoupenko2010cef, Wang2010utb}, which is expected to fix the 3BP. For intermediate cases, the problem gets very complex. Here, effective-range effects have been shown to cause corrections to the Efimov resonance positions and to the connections between different features \cite{Thogersen2008upo, Chen2011ttb}. For open-channel dominated resonances, the van der Waals length $R_{\rm vdW}$ or alternatively the mean scattering length $\bar{a} = 0.956\,R_{\rm vdW}$ (Sec.\ \ref{basic}) appear as the only additional length scale. This suggests a fixed relation between $a^{(0)}_-$ and $\bar{a}$. In fact, our actual Cs value $a^{(0)}_-/\bar{a} = -9.5(4)$ is remarkably close to corresponding values for $^7$Li \cite{Gross2010nsi,Pollack2009uit} and $^6$Li \cite{Ottenstein2008cso,Huckans2009tbr, Williams2009efa,Wenz2009uti}, which vary in the range between $-8$ and $-10$. This is an exciting hint, which motivates further investigations both theoretically and experimentally.

\begin{figure*}[t]
\centering
\includegraphics[width=.7\textwidth] {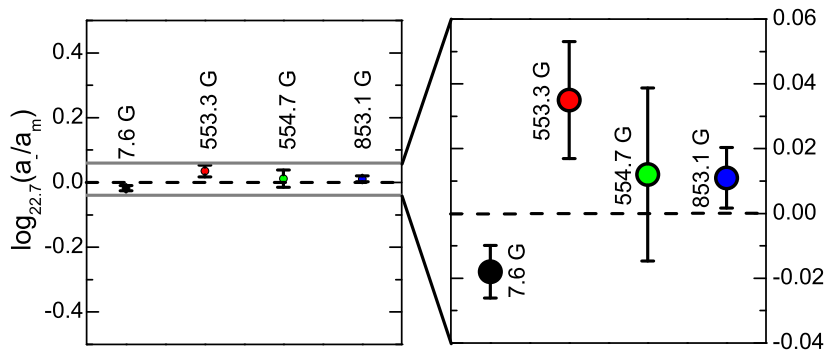}
\caption{Position of the three-body loss features of cesium. The dashed line corresponds to the weighted mean value of  $-921\,a_0$. While in the left panel the logarithmic scale (to the basis of 22.7) covers one Efimov period, only one tenth of that period (factor of $22.7^{1/10} = 1.37$) is shown in the right panel. The results obtained at 498.1 and 819.89\,G are not shown here since the conversion $a(B)$ is subject to systematic errors, which are not fully under control; see discussion in Sec.\,\ref{sec:gEfi}.}
\label{csefimov}
\end{figure*}

To clarify the issue of the 3BP, investigations of Efimov features at different Feshbach resonances in the same atomic system provide interesting information. In Ref.~\cite{Gross2010nsi} the Efimov scenario was investigated in two channels that differed in the nuclear spin orientation. The comparison showed no significant difference in the 3BP, thus suggesting at least its independence of the nuclear spin. In Ref.~\cite{Berninger2011uot} we have determined the 3BP for four different triatomic Efimov resonances in Cs; see also Sec.~\ref{sec:5}. The results are illustrated in Fig.\,\ref{csefimov}. The left panel shows the obtained values on a logarithmic scale that covers the full Efimov period (factor of 22.7); the right panel shows a zoom in covering one tenth of the Efimov period. The results show only very small variations around an average value of $a^{(0)}_- = -921\,a_0$, staying within a few percent of the Efimov period. The deviations from the average value are not statistically significant\footnote{Between the values determined for the two broad resonances at 7.6 and 853\,G we find a possible small aberration of about $2.5$ standard deviations. This may be accidental but it may also hint at a small change in the 3BP.}. Thus our data are consistent with a constant or weakly varying 3BP. This observation in Cs rules out a scenario where the 3BP is essentially random for each new resonance \cite{Dincao2009tsr}.

It is interesting to note that recent experimental results \cite{Nakajima2011moa} on atom-dimer interactions in three-component spin mixtures of $^6$Li indicate small variations of the 3BP, although all observations on three-atom interactions in the same system are consistent with a constant 3BP. Our observations on Cs, where the three-body recombination minimum follows universal relations (Sec. \ref{sec:5}) while the atom-dimer resonance (Sec. \ref{sec:6}) shows substantial deviations, point into a similar direction. This is another unresolved issue regarding the 3BP.
  Let us finally discuss the decay parameter $\eta_-$ (or $\eta_+$ for $a>0$). This quantity is sometimes referred to as the imaginary part of the 3BP, and it is related to the lifetime of an Efimov trimer against spontaneous decay into a more deeply bound dimer state and a free atom. This process sensitively depends on the particular spectrum of weakly-bound dimer states \cite{Wenz2009uti, Hammer2007erc}. Thus the decay parameter is expected to be a non-universal quantity.

\section{Universal four-body states tied to an Efimov trimer}
 	  \label{sec:9}

The Efimov effect shows that a universal sequence of three-body bound states exists for resonant two-body interaction. One could easily be  tempted to extend this scenario to $N$-body bound states and resonant ($N-1$)-body interactions. However, soon after Efimov's original  work, it was demonstrated that no ``true'' Efimov effect can exist for four or more identical particles \cite{Amado1973tin}, underlining the distinctive and exclusive character of the effect to three-body systems. In fact to have a genuine Efimovian character, an $N$-body state should not only exhibit a precise scaling invariance but also a well-defined behavior at threshold, meaning that an $(N-1)$-body bound state should cause the appearance of an infinite ladder of $N$-body states at the $N$-body atomic threshold. Nevertheless, the question of possible universal tetramer states remained an open issue. It was later pointed out that universal four-body states, which are not genuinely Efimovian but still Efimov-related, or even true four-body Efimov states in light-heavy particle systems might exist \cite{Adhikari1981fbe,Naus1987tee,Sorensen2002ctb,Platter2004fbs,Yamashita2006fbs,Hanna2006eas,Thogersen2008nbe,Castin2010fbe}.

\begin{figure}[t]
\centering
\includegraphics[width=0.75\textwidth] {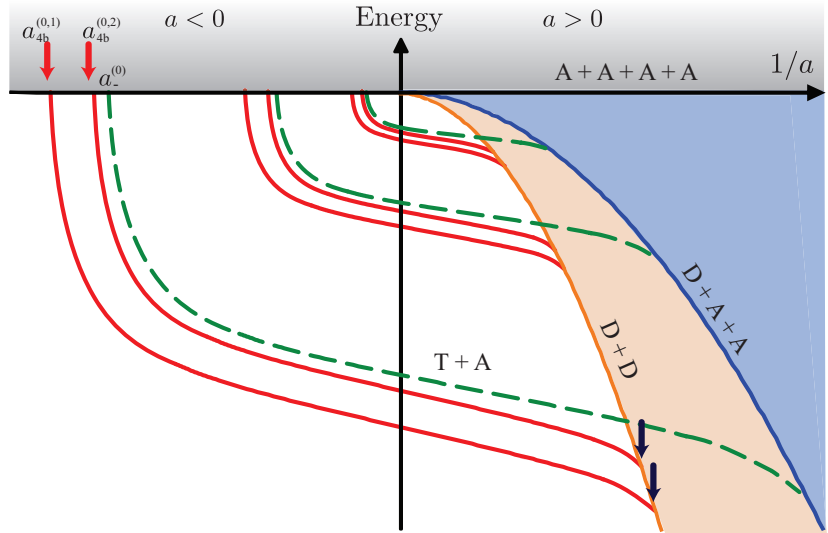}
\caption{Extension of Efimov's scenario to four identical bosons. The solid lines  illustrate the pairs of universal tetramer states linked to each Efimov trimer. In the four-body picture, the Efimov trimers correspond to the trimer-atom thresholds (dashed lines). A four-body state crosses the four-atom threshold (A+A+A+A) and merges with the dimer-dimer threshold (D+D). The arrows indicate the positions where the energetically lowest pair of four-body states couples to the atomic threshold and to the dimer-dimer threshold.}
\label{fig:four}
\end{figure}

Two theory groups, one in Bonn/Ohio \cite{Hammer2007upo} and the other one at JILA in Boulder \cite{Vonstecher2009sou}, have made a fundamental step in understanding the four-body problem for identical bosons. They predicted the existence of
a small finite number of universal four-body bound states tied to each Efimov trimer. As suggested in Ref.\,\cite{Hammer2007upo} and more rigorously demonstrated in Ref.\,\cite{Vonstecher2009sou}, there are exactly two four-body states. The key point  is that in the proximity of an Efimov trimer the four-body potential is strong enough to support  two, but only two bound states. The extended Efimov scenario for four particles is illustrated in Fig.\,\ref{fig:four}. In Ref.\,\cite{Vonstecher2009sou}, the universal scaling factors between an Efimov trimer and the pair of tetramer states have been calculated. The two tetramer states cross the four-atom threshold at  $a_{\rm 4b}^{(n,1)}=0.43 a_-^{(n)}$ and $a_{\rm 4b}^{(n,2)}=0.90 a_-^{(n)}$ \cite{Vonstecher2009sou} and merge with the dimer-dimer threshold at $2.37 a_*^{(n)}$ and $6.6 a_*^{(n)}$ \cite{Dincao2009ufb}, respectively.

\begin{figure}[t]
\centering
\includegraphics[width=0.75\textwidth] {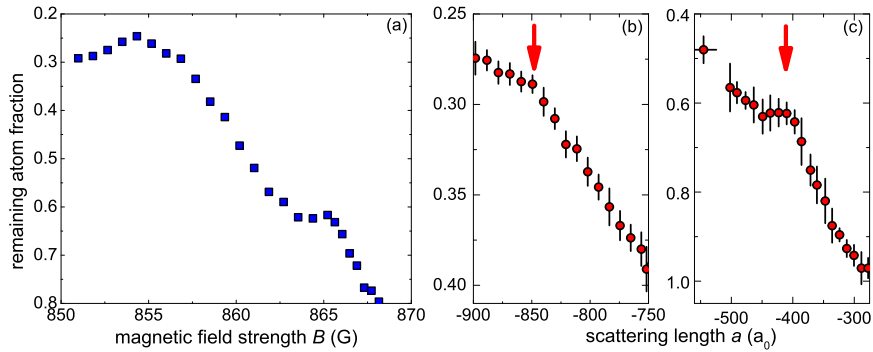}
\caption{ Observation of four-body resonances at high magnetic fields in Cs. (a)  Remaining number of atoms as a function of magnetic field strength $B$. The measurements are performed on a sample of  $6 \times 10^4$ atoms at $100$~nK. The waiting time in the optical trap is $2$\,s. (b)-(c) Remaining atom number as a function of the scattering length  in the vicinity of the two four-body resonances.  The two four-body loss resonances are indicated by the arrows. The measurements are performed with optimized parameters: (b) $5 \times 10^4$ atoms at $70$~nK with a waiting time of $0.01$\,s, and (c) $2.5 \times 10^4$ atoms at $50$~nK with a waiting time of $1$\,s. }
\label{fig:fourloss}
\end{figure}

Remarkably, collisional rates can still serve as central observable in experiments. In general, four-body collisions are negligible in ultracold gases since this high-order process is usually completely masked by dominant three-body mechanisms. However, when a tetramer state crosses the atomic threshold the recombination rate resonantly increases and becomes sizeable \cite{Mehta2009gtd}. In Ref.\,\cite{Vonstecher2009sou}, the authors suggested that already our early 2006 investigations \cite{Kraemer2006efe} showed indications of the above-described four-body states. In a later dedicated experiment, we carefully investigated this issue in the low-field region \cite{Ferlaino2009efu}.
In the vicinity of the Efimov trimer, we observed a pair of four-body resonances at the four-atom threshold and verified the  predicted universal relations between the three- and four-body states \cite{Vonstecher2009sou}. Our findings are based on four-body recombination rate measurements and  loss spectroscopy in optical trapped samples of Cs atoms. In particular we could clearly obsverve a four-body behavior in atom-number decay measurements and extract the four-body relaxation rate coefficient $L_4$, accordingly to the rate equation $\dot{n}_{\rm A}=-L_3n_{\rm A}^3-L_4n_{\rm A}^4$, where $n_A$ is the atomic density. Evidence for the existence of the four-body states has also been  obtained with $^{39}$K in the Florence experiments \cite{Zaccanti2009ooa} and with $^7$Li in the Rice experiments \cite{Pollack2009uit}.

We now report on novel observations of the four-body states in the high-field region. We followed a strategy similar to the one used in the low-field region. We prepared an ultracold sample in a crossed dipole trap and we performed loss spectroscopy as a function of the magnetic field strength for a fixed storing time in the trap. Our results are shown in Fig.\,\ref{fig:fourloss}. We first performed a magnetic-field scan over a comparatively wide range of values and we identified the two four-body resonances (Fig.\,\ref{fig:fourloss}(a)), we then focused on the two regions of interest and we repeated the measurements using optimized time sequences (Fig.\,\ref{fig:fourloss}(b-c)).
The two four-body resonances appear at $865.4(5)~\rm{G}$ and $855.0(2)~\rm{G}$, respectively. Using the previously discussed $a({\rm B})$ conversion, we find   $a_{\rm 4b}^{(0,1)}=-444(8)\,a_0$ and $a_{\rm 4b}^{(0,2)}=-862(9)\,a_0$, respectively. These values are remarkably close to the ones found in the low-field region \cite{Ferlaino2009efu}. In both the low- and high-field region our observations confirm the universal relation predicted in \cite{Vonstecher2009sou}. In particular, we found $a_{\rm 4b}^{(0,1)}/a_-^{(0)}=0.47(1)$, $a_{\rm 4b}^{(0,2)}/a_-^{(0)}=0.87(1)$, and $a_{\rm 4b}^{(0,1)}/a_-^{(0)}=0.46(2)$ and $a_{\rm 4b}^{(0,2)}/a_-^{(0)}=0.91(3)$ at low- and high-field, respectively.

Our results confirm once more the robustness of universality through different Feshbach resonances and magnetic fields and demonstrate the ability to observe higher-order few-body phenomena with ultracold gases. Recent theoretical investigations have predicted a further extension of the Efimov scenario, where 5, 6, 7 or more bodies can bind even if no bound subsystem exist  \cite{Hanna2006eas,Vonstecher2010wbc,vonStecher2011ufa}. Cluster with up to $N=40$ atoms have been investigated and the scattering-length values at which successive $N$-bosons systems cross the corresponding atomic threshold have been calculated. This exciting direction of few-body physics awaits experimental confirmation.

\section{Few-body phenomena in other ultracold quantum gas experiments}
  	 \label{sec:10others}
	
In the last few years, Efimov states and related universal features have been observed in various other ultracold systems \cite{Ottenstein2008cso,Huckans2009tbr,Zaccanti2009ooa,Barontini2009ooh,Gross2009oou,
Pollack2009uit,Nakajima2010nea, Gross2010nsi,Lompe2010ads,Nakajima2011moa}, extending and complementing the Cs observations discussed above. The experiments confirmed basic elements of the Efimov scenario, but also triggered new questions on universality and its connection to real-world systems. Here we briefly review the main findings of the experiments performed by other groups in various quantum gases with tunable interactions.

An important breakthrough for the three-boson system, reported in Refs.~\cite{Zaccanti2009ooa,Pollack2009uit}, was to observe the scaling factor between the Efimov features by detecting consecutive loss resonances in recombination experiments. Other experiments concentrated on the extension of the Efimov scenario to non-identical particles, using a spin mixture of fermionic atoms \cite{Ottenstein2008cso,Huckans2009tbr,Nakajima2010nea} or a mixture of different atomic species \cite{Barontini2009ooh}.  A different approach to test the robustness of universality was based on the observation of Efimov features on Feshbach resonances occurring in different spin channels \cite{Gross2009oou,Gross2010nsi}, on different Feshbach resonances  in the same entrance channel \cite{Berninger2011uot}, and by measuring the Efimov trimer binding energy \cite{Lompe2010ads, Nakajima2011moa}.

Experiments on $^{39}$K  performed in Florence provided a first confirmation of the Efimov's universal scaling factor  \cite{Zaccanti2009ooa}. In the positive $a$ region, two consecutive recombination minima were observed. The ratio between the positions of the minima was found to be $a^{(1)}_+/a^{(0)}_+=25(4)$, which agrees within the uncertainty with the universal scaling factor of 22.7; see Table \ref{tab:1}. In addition, two loss resonances  were observed in the atomic samples at positive scattering length values.
These loss resonances in atomic collisions were interpreted as being caused by the underlying atom-dimer Efimov resonances through multiple collision processes.
In the negative $a$ region, a single tri-atomic resonance was revealed. Remarkably, all the measured ratios connecting features at positive and negative $a$  showed a  substantial deviation of up to about $60\%$ from the predicted universal ratios.
In particular, the ratio between the tri-atomic and atom-dimer Efimov resonance position $a^{(1)}_{*}/|a^{(0)}_{-}|$ resulted to be about $50\%$ smaller than the expected one.  A similar reduction was also observed in our Cs experiments \cite{Knoop2009ooa}, as discussed in Sec.\,\ref{sec:6}.

The Rice group \cite{Pollack2009uit} and the Bar Ilan group \cite{Gross2009oou,Gross2010nsi} studied the Efimov spectrum and the corresponding universal relations in ultracold gases of $^7$Li. In Ref.\,\cite{Pollack2009uit}, the authors reported a number of recombination features for both positive and negative $a$ values that they interpreted as being caused by  consecutive tri-atomic and atom-dimer Efimov resonances, and recombination minima. However the interpretation of the loss features in terms of the Efimov scenario partially remains an open issue because of discrepancies in the $a(B)$ conversion used by both groups \cite{Pollack2009uit,Gross2010nsi}  and further investigations are need to unambiguously assign the observed features. In Ref.\,\cite{Gross2009oou,Gross2010nsi}, the Bar Ilan group performed recombination-rate measurements with atoms in two spin channels and compared their results. They observed  a tri-atomic Efimov resonance appearing for both spin states at the same value of the scattering length $a_-(0)$. In addition, they found the corresponding recombination minimum at a position $a_+(0)$, which is in an excellent agreement with universal theory. Combining the Bar Ilan with our results, we can conclude that the Efimov resonances are largely insensitive to the entrance spin channel \cite{Gross2009oou} and to the character of the closed channel underlying the Feshbach resonance used to tune the interactions into the resonant regime \cite{Berninger2011uot}.

The Efimov effect does not only occur in systems of three identical bosons as we have discussed so far, but can be generalized to the case of distinguishable particles. In ultracold gas experiments, a system of non-identical atoms can be realized using fermions or bosons  in spin-state mixtures  or heteronuclear mixtures of two or more atomic species.

An intriguing case is provided by three-component spin mixture in ultracold Fermi gases of $^6$Li atoms, where the atoms are equally distributed in the three lowest spin channels.
A remarkable feature of this system is  that each of the three relevant spin channels exhibits a broad $s$-wave Feshbach resonance \cite{Bartenstein2005pdo}; all three resonances overlap in a wide magnetic field range. This situation provides a very general realization of the Efimov scenario where three non-identical particles resonantly interact with each other, with all three scattering lengths, corresponding to the three two-body sub-systems, being very large.
Three groups, one in Heidelberg, one at Penn State University, and one in Tokyo have used this system to explore Efimov physics \cite{Wenz2009uti,Ottenstein2008cso,Huckans2009tbr,Nakajima2010nea,Lompe2010rfa,Nakajima2011moa}. Tri-atomic Efimov resonances were observed in recombination-rate experiments \cite{Wenz2009uti,Ottenstein2008cso,Huckans2009tbr} and the  binding energies of an Efimov trimer state were measured as a function of the scattering length via radio-frequency spectroscopy measurements \cite{Lompe2010rfa, Nakajima2011moa}.

The Efimov problem of distinguishable particles can also be realized with heteronuclear mixtures. Remarkably, the universal scaling factor strongly depends on  the mass ratio and  very dense spectra can appear for mass imbalanced systems \cite{Dincao2006eto,Dincao2006mdo}. For instance, the scaling factor reduces from 22.7 to about 5 passing from a system of  three identical bosons to the Yb-Yb-Li system, which has a mass imbalance factor of about 28. So far, Efimov states in heteronuclear mixtures were observed only in $^{41}$K-$^{87}$Rb mixtures by the Florence group \cite{Barontini2009ooh}. They observed two tri-atomic Efimov resonances, one connected to the K-Rb-Rb system and the other one to the K-K-Rb combination, that showed very different collisional rate values and decay parameter $\eta_-$, pointing to the difference in behavior between the light-heavy-heavy and the light-light-heavy three-body system.

Thanks to all these experiments on ultracold gases,  the Efimov scenario could be firmly established and important steps beyond the three-boson systems were made by investigating universal four-body physics. However, the experimental observations also raised questions on how far universal theories can be applied to describe real systems and to which extent non-universal and short-range physics affect Efimov's original scenario.

\section{Conclusions and outlook}
  	 \label{sec:conclu}

Efimov's scenario is now well established as the paradigm of universal few-body physics. The experimental possibilities opened up by ultracold gases with tunable interactions have given an enormous boost to research on few-body phenomena. With the recent experiments, the basic phenomenology and many key elements of universal few-body physics could be confirmed, but there are still open issues awaiting for clarifications. The latter mainly concerns the question in how far real-world systems follow the idealized universal behavior or at which point nonuniversal physics takes over.

In a more general sense, an increasing number of few-body phenomena involving three or more resonantly interacting particles are being discovered in a great variety of different situations and environments; see e.g.\ \cite{Nishida2008ufg, Nishida2009cie, Levinsen2009ads, Castin2010fbe, Vonstecher2010wbc, Wang2011eef}. Here, the properties of the particles and their interactions are crucial for the particular nature of a few-body phenomenon. The composition of different masses, the quantum statistics, the interactions via $s$- and higher partial waves, possible dipolar interactions, and the external trapping environment offer a huge parameter space for a wealth of phenomena to occur.

An intriguing general question is how few-body phenomena affect the many-body physics of a strongly interacting quantum system or, in other words, how one can better understand the properties of a many-body system based on few-body physics. On one hand, we have to understand loss and recombination properties in such a system, which in many cases impose limitations to the experiments. On the other hand, the few-body approach can give new insights into the many-body physics of a strongly interacting system. In the context of Fermi gases, important examples are given by the stability of dimers \cite{Petrov2004wbm}, the few-body perspective on the BEC-BCS crossover problem \cite{Vonstecher2007sad}, and the virial expansion \cite{Liu2009vef}. Moreover, recent theoretical work has shown that general connections exist between few-body interactions and the essential properties of many-body systems \cite{Tan2009eoa, Braaten2009htt}.

Many additional possibilities result from the control of the external degrees of freedom, offered in a unique way by ultracold gases in an optical trapping environment. Optical lattices allow for tight confinement in one, two, or three dimensions allowing for the realization of low-dimensional systems with highly non-trivial properties. Such lattices also mimic the period environment of a solid-state crystal. A new avenue of research is to exploit few-body interactions to introduce higher-order correlations into such systems \cite{Daley2009atb}, which in a controlled way can lead to many-body phases with novel properties. Research along these lines is in an early stage, but may have strong implications for many-body physics in optical lattices.

Forty years after its prediction, Efimov physics stands for a new research field with many intriguing opportunities. Few-body phenomena are ubiquitous in strongly interacting particle systems, and learning more about the general nature and particular properties will enable us to better understand and control the physics of quantum matter under conditions of strong interactions.

\section{Acknowledgements \& Supports}

This work was supported by the Austrian Science Fund FWF within project P23106. A.Z.\ is supported within the Marie Curie Project No.\ 254987 (LatTriCs) of the European Commission.

%


\bibliographystyle{ieeetr}

\begin{thebibliography}{10}

\bibitem{Efimov1970ela}
V.~Efimov, ``Energy levels arising from resonant two-body forces in a
  three-body system,'' {\em Phys. Lett. B}, vol.~33, 563--564, 1970.

\bibitem{Kraemer2006efe}
T.~Kraemer, M.~Mark, P.~Waldburger, J.~G. Danzl, C.~Chin, B.~Engeser, A.~D.
  Lange, K.~Pilch, A.~Jaakkola, H.-C. N\"agerl, and R.~Grimm, ``Evidence for
  Efimov quantum states in an ultracold gas of cesium atoms,'' {\em Nature},
  vol.~440, 315--318, 2006.

\bibitem{Ottenstein2008cso}
T.~B. Ottenstein, T.~Lompe, M.~Kohnen, A.~N. Wenz, and S.~Jochim, ``Collisional
  stability of a three-component degenerate Fermi gas,'' {\em Phys. Rev.
  Lett.}, vol.~101,  203202, 2008.

\bibitem{Huckans2009tbr}
J.~H. Huckans, J.~R. Williams, E.~L. Hazlett, R.~W. Stites, and K.~M. O'Hara,
  ``Three-body recombination in a three-state fermigas with widely tunable
  interactions,'' {\em Phys. Rev. Lett.}, vol.~102,  165302, 2009.

\bibitem{Knoop2009ooa}
S.~Knoop, F.~Ferlaino, M.~Mark, M.~Berninger, H.~Sch\"obel, H.-C. N\"agerl, and
  R.~Grimm, ``Observation of an Efimov-like trimer resonance in ultracold
  atom-dimer scattering,'' {\em Nature Phys.}, vol.~5, 227--230, 2009.

\bibitem{Zaccanti2009ooa}
M.~Zaccanti, B.~Deissler, C.~D'Errico, M.~Fattori, M.~Jona-Lasinio,
  S.~M\"{u}ller, G.~Roati, M.~Inguscio, and G.~Modugno, ``Observation of an
  Efimov spectrum in an atomic system,'' {\em Nature Phys.}, vol.~5,  586,
  2009.

\bibitem{Barontini2009ooh}
G.~Barontini, C.~Weber, F.~Rabatti, J.~Catani, G.~Thalhammer, M.~Inguscio, and
  F.~Minardi, ``Observation of heteronuclear atomic Efimov resonances,'' {\em
  Phys. Rev. Lett.}, vol.~103,  043201, 2009.

\bibitem{Gross2009oou}
N.~Gross, Z.~Shotan, S.~Kokkelmans, and L.~Khaykovich, ``Observation of
  universality in ultracold $^7$Li three-body recombination,'' {\em Phys. Rev.
  Lett.}, vol.~103,  163202, 2009.

\bibitem{Pollack2009uit}
S.~E. Pollack, D.~Dries, and R.~G. Hulet, ``Universality in three- and
  four-body bound states of ultracold atoms,'' {\em Science}, vol.~326,
  1683--1686, 2009.

\bibitem{Nakajima2010nea}
S.~Nakajima, M.~Horikoshi, T.~Mukaiyama, P.~Naidon, and M.~Ueda, ``Nonuniversal
  Efimov atom-dimer resonances in a three-component mixture of $^{6}$Li,'' {\em
  Phys. Rev. Lett}, vol.~105,  023201, 2010.

\bibitem{Gross2010nsi}
N.~Gross, Z.~Shotan, S.~Kokkelmans, and L.~Khaykovich,
  ``Nuclear-spin-independent short-range three-body physics in ultracold
  atoms,'' {\em Phys. Rev. Lett}, vol.~105,  103203, 2010.

\bibitem{Lompe2010ads}
T.~Lompe, T.~B. Ottenstein, F.~Serwane, K.~Viering, A.~N. Wenz, G.~Z\"{u}rn,
  and S.~Jochim, ``Atom-dimer scattering in a three-component Fermi gas,'' {\em
  Phys. Rev. Lett}, vol.~105,  103201, 2010.

\bibitem{Nakajima2011moa}
S.~Nakajima, M.~Horikoshi, T.~Mukaiyama, P.~Naidon, and M.~Ueda, ``Measurement
  of an Efimov trimer binding energy in a three-component mixture of
  $^{6}$Li,'' {\em Phys. Rev. Lett.}, vol.~106,  143201, 2011.

\bibitem{Efimov2009gtt}
V.~Efimov, ``Giant trimer true to scale,'' {\em Nature Phys.}, vol.~5,
  533--534, 2009.

\bibitem{Greene2010uif}
C.~H. Greene, ``Universal insights from few-body land,'' {\em Phys. Today},
  vol.~63(3), 40--45, 2010.

\bibitem{Ferlaino2010fyo}
F.~Ferlaino and R.~Grimm, ``Forty years of Efimov physics: How a bizarre
  prediction turned into a hot topic,'' {\em Physics}, vol.~3,  9, 2010.

\bibitem{Efimov1979lep}
V.~Efimov, ``Low-energy properties of three resonantly interacting particles,''
  {\em Sov. J. Nuc. Phys.}, vol.~29, 546--553, 1979.

\bibitem{Ferlaino2009efu}
F.~Ferlaino, S.~Knoop, M.~Berninger, W.~Harm, J.~P. {D'Incao}, H.-C. N\"agerl,
  and R.~Grimm, ``Evidence for universal four-body states tied to an Efimov
  trimer,'' {\em Phys. Rev. Lett.}, vol.~102,  140401, 2009.

\bibitem{Berninger2011uot}
M.~Berninger, A.~Zenesini, B.~Huang, W.~Harm, H.-C. N\"{a}gerl, F.~Ferlaino,
  R.~Grimm, P.~S. Julienne, and J.~M. Hutson, ``Universality of the three-body
  parameter for Efimov states in ultracold cesium,'' {\em Phys. Rev. Lett.}, vol.~107, 120401, 2011.

\bibitem{Taylor1983stt}
J.~R. Taylor, {\em Scattering Theory: The Quantum Theory of Nonrelativistic
  Collisions}.
\newblock Dover Books on Engineering, 1983.

\bibitem{Chin2010fri}
C.~Chin, R.~Grimm, P.~S. Julienne, and E.~Tiesinga, ``Feshbach resonances in
  ultracold gases,'' {\em Rev. Mod. Phys.}, vol.~82, 1225--1286, 2010.

\bibitem{Gao1998qdt}
B.~Gao, ``Quantum-defect theory of atomic collisions and molecular vibration
  spectra,'' {\em Phys. Rev. A}, vol.~58, 4222--4225, 1998.

\bibitem{Braaten2006uif}
E.~Braaten and H.-W. Hammer, ``Universality in few-body systems with large
  scattering length,'' {\em Phys. Rep.}, vol.~428, 259--390, 2006.

\bibitem{Derevianko1999hpc}
A.~Derevianko, W.~R. Johnson, M.~S. Safronova, and J.~F. Babb, ``High-precision
  calculations of dispersion coefficients, static dipole polarizabilities, and
  atom-wall interaction constants for alkali-metal atoms,'' {\em Phys. Rev.
  Lett.}, vol.~82, 3589--3592, 1999.

\bibitem{Jensen2004sar}
A.~S. Jensen, K.~Riisager, D.~V. Fedorov, and E.~Garrido, ``Structure and
  reactions of quantum halos,'' {\em Rev. Mod. Phys.}, vol.~76, 215--261,
  2004.

\bibitem{Gribakin1993csl}
G.~F. Gribakin and V.~V. Flambaum, ``Calculation of the scattering length in
  atomic collisions using the semiclassical approximation,'' {\em Phys. Rev.
  A}, vol.~48, 546--553, 1993.

\bibitem{Petrov2004tbp}
D.~S. Petrov, ``Three-boson problem near a narrow Feshbach resonance,'' {\em
  Phys. Rev. Lett.}, vol.~93,  143201, 2004.

\bibitem{Chin2000hrf}
C.~Chin, V.~Vuleti\'{c}, A.~J. Kerman, and S.~Chu, ``High resolution Feshbach
  spectroscopy of cesium,'' {\em Phys. Rev. Lett.}, vol.~85,  2717, 2000.

\bibitem{Chin2004pfs}
C.~Chin, V.~Vuleti\'c, A.~J. Kerman, S.~Chu, E.~Tiesinga, P.~J. Leo, and C.~J.
  Williams, ``Precision Feshbach spectroscopy of ultracold Cs$_2$,'' {\em Phys.
  Rev. A}, vol.~70,  032701, 2004.

\bibitem{Mark2007sou}
M.~Mark, F.~Ferlaino, S.~Knoop, J.~G. Danzl, T.~Kraemer, C.~Chin, H.-C.
  N\"{a}gerl, and R.~Grimm, ``Spectroscopy of ultracold trapped cesium Feshbach
  molecules,'' {\em Phys. Rev. A}, vol.~76,  042514, 2007.

\bibitem{Berninger2011hmf}
M.~Berninger, A.~Zenesini, B.~Huang, H.-C. N\"{a}gerl, F.~Ferlaino, R.~Grimm,
  P.~S. Julienne, and J.~M. Hutson., ``High magnetic-field scattering
  properties of ultracold cs atoms,'' {\em In preparation}, 2011.

\bibitem{Leo2000cpo}
P.~J. Leo, C.~J. Williams, and P.~S. Julienne, ``Collision properties of
  ultracold $^{133}${C}s atoms,'' {\em Phys. Rev. Lett.}, vol.~85,
  2721--2724, 2000.

\bibitem{Zenesini2011CBE}
A.~Zenesini, M.~Berninger, B.~Huang, H.-C. N\"{a}gerl, F.~Ferlaino, and
  R.~Grimm., ``Creation of Bose Einstein condensates of cesium at high magnetic
  fields,'' {\em In preparation}, 2011.

\bibitem{Lee2007ete}
M.~D. Lee, T.~K\"{o}hler, and P.~S. Julienne, ``Excited Thomas-Efimov levels in
  ultracold gases,'' {\em Phys. Rev. A}, vol.~76, 012720, 2007.

\bibitem{Chin2001hpf}
C.~Chin, V.~Vuleti\'{c}, A.~J. Kerman, and S.~Chu, ``High precision Feshbach
  spectroscopy of ultracold cesium collisions,'' {\em Nucl. Phys. A}, vol.~684,
  641C--645C, 2001.

\bibitem{Gustavsson2008coi}
M.~Gustavsson, E.~Haller, M.~J. Mark, J.~G. Danzl, G.~Rojas-Kopeinig, and H.-C.
  N\"agerl, ``Control of interaction-induced dephasing of Bloch oscillations,''
  {\em Phys. Rev. Lett.}, vol.~100,  080404, 2008.

\bibitem{Gustavsson2008PhD}
M.~Gustavsson, {\em A quantum gas with tunable interactions in an optical
  lattice}.
\newblock PhD thesis, University of Innsbruck, 2008.

\bibitem{Nielsen1999ler}
E.~Nielsen and J.~H. Macek, ``Low-energy recombination of identical bosons by
  three-body collisions,'' {\em Phys. Rev. Lett.}, vol.~83, 1566--1569,
  1999.

\bibitem{Esry1999rot}
B.~D. Esry, C.~H. Greene, and J.~P. Burke, ``Recombination of three atoms in
  the ultracold limit,'' {\em Phys. Rev. Lett.}, vol.~83, 1751--1754, 1999.

\bibitem{Weber2003tbr}
T.~Weber, J.~Herbig, M.~Mark, H.-C. N\"agerl, and R.~Grimm, ``Three-body
  recombination at large scattering lengths in an ultracold atomic gas,'' {\em
  Phys. Rev. Lett.}, vol.~91, 123201, 2003.

\bibitem{Fedichev1996tbr}
P.~O. Fedichev, M.~W. Reynolds, and G.~V. Shlyapnikov, ``Three-body
  recombination of ultracold atoms to a weakly bound $s$ level,'' {\em Phys.
  Rev. Lett.}, vol.~77, 2921--2924, 1996.

\bibitem{Wenz2009uti}
A.~N. Wenz, T.~Lompe, T.~B. Ottenstein, F.~Serwane, G.~Z\"urn, and S.~Jochim,
  ``Universal trimer in a three-component Fermi gas,'' {\em Phys. Rev. A},
  vol.~80,  040702(R), 2009.

\bibitem{Dincao2004lou}
J.~P. D'Incao, H.~Suno, and B.~D. Esry, ``Limits on universality in ultracold
  three-boson recombination,'' {\em Phys. Rev. Lett.}, vol.~93,  123201, 
  2004.

\bibitem{Braaten2004edr}
E.~Braaten and H.~W. Hammer, ``Enhanced dimer relaxation in an atomic and
  molecular Bose-Einstein condensate,'' {\em Phys. Rev. A}, vol.~70,  042706,
  2004.

\bibitem{DIncao2005sls}
J.~P. D'Incao and B.~D. Esry, ``Scattering length scaling laws for ultracold
  three-body collisions,'' {\em Phys. Rev. Lett.}, vol.~94,  213201, 2005.

\bibitem{Braaten2007rdr}
E.~Braaten and H.-W. Hammer, ``Resonant dimer relaxation in cold atoms with a
  large scattering length,'' {\em Phys. Rev. A}, vol.~75,  052710,
  2007.

\bibitem{Weber2003bec}
T.~Weber, J.~Herbig, M.~Mark, H.-C. N\"agerl, and R.~Grimm, ``{Bose-Einstein}
  condensation of cesium,'' {\em Science}, vol.~299, 232--235, 2003.

\bibitem{Kraemer2004opo}
T.~Kraemer, J.~Herbig, M.~Mark, T.~Weber, C.~Chin, H.-C. N\"agerl, and
  R.~Grimm, ``Optimized production of a cesium Bose-Einstein condensate,'' {\em
  Appl. Phys. B}, vol.~79, 1013--1019, 2004.

\bibitem{Rychtarik2004tdb}
D.~Rychtarik, B.~Engeser, H.-C. N\"agerl, and R.~Grimm, ``Two-dimensional
  Bose-Einstein condensate in an optical surface trap,'' {\em Phys. Rev.
  Lett.}, vol.~92, 173003, 2004.

\bibitem{Naegerl2006eef}
H.-C. N\"agerl, T.~Kraemer, M.~Mark, P.~Waldburger, J.~G. Danzl, C.~Chin,
  B.~Engeser, A.~D. Lange, K.~Pilch, A.~Jaakkola, and R.~Grimm, ``Experimental
  evidence for Efimov quantum states,'' {\em Atomic Physics 20 AIP Conf.
  Proc.}, vol.~869, 269--277, 2006.

\bibitem{Bringas2004tcr}
F.~Bringas, M.~T. Yamashita, and T.~Frederico, ``Triatomic continuum resonances
  for large negative scattering lengths,'' {\em Phys. Rev. A}, vol.~69,
   040702, 2004.

\bibitem{Yamashita2007tbr}
M.~Yamashita, T.~Frederico, and L.~Tomio, ``Three-boson recombination at
  ultralow temperatures,'' {\em Physics Letters A}, vol.~363, 468
  -- 472, 2007.

\bibitem{Dincao2009tsr}
J.~P. D'Incao, C.~H. Greene, and B.~D. Esry, ``The short-range three-body phase
  and other issues impacting the observation of Efimov physics in ultracold
  quantum gases,'' {\em J. Phys. B: At. Mol. Opt. Phys.}, vol.~42,  044016,
  2009.

\bibitem{Platter2008sfa}
L.~Platter and J.~R. Shepard, ``Scaling functions applied to three-body
  recombination of $^133$Cs atoms,'' {\em Phys. Rev. A}, vol.~78,  062717,
  Dec 2008.

\bibitem{Massignan2008esn}
P.~Massignan and H.~T.~C. Stoof, ``Efimov states near a Feshbach resonance,''
  {\em Phys. Rev. A}, vol.~78,  030701, 2008.

\bibitem{Jonalasinio2010tru}
M.~Jona-Lasinio and L.~Pricoupenko, ``Three resonant ultracold bosons:
  Off-resonance effects,'' {\em Phys. Rev. Lett.}, vol.~104,  023201,
  2010.

\bibitem{Ferlaino2009ufm}
F.~Ferlaino, S.~Knoop, and R.~Grimm, {\em Cold Molecules: Theory, Experiment,
  Applications}, chapter in ``Ultracold Feshbach molecules'.
\newblock edt. by R. V. Krems, B. Friedrich, and W. C. Stwalley (CRC Press, Boca Raton, 2009).

\bibitem{Knoop2009ooab}
S. ~Knoop, F.~Ferlaino, M.~Berninger, M.~Mark, H.-C.~N\"agerl, R.~Grimm, ``Observation of an Efimov resonance in an ultracold mixture of atoms and weakly bound dimers,'' {\em J. Phys.: Conf. Ser.} 194, 012064 (2009), arXiv:0907.4510.

\bibitem{Braaten2009erd}
E.~Braaten and H.-W. Hammer, ``Erratum: Resonant dimer relaxation in cold atoms
  with a large scattering length,'' {\em Phys.
  Rev. A}, vol.~79,  039905, 2009.

\bibitem{Helfrich2009rad}
K.~Helfrich and H.~W. Hammer, ``Resonant atom-dimer relaxation in ultracold
  atoms,'' {\em Europhys. Lett.}, vol.~86,  53003, 2009.

\bibitem{Esry2011pc}
B.~D. Esry, ``private comunication,'' 2011.

\bibitem{Thomas1935tib}
L.~H. Thomas, ``The interaction between a neutron and a proton and the
  structure of H$^3$,'' {\em Phys. Rev.}, vol.~47, 903--909, 1935.

\bibitem{Bedaque1999rof}
P.~F. Bedaque, H.-W. Hammer, and U.~van Kolck, ``Renormalization of the
  three-body system with short-range interactions,'' {\em Phys. Rev. Lett.},
  vol.~82, 463--467, 1999.

\bibitem{Soldan2003tbn}
P.~Sold\'an, M.~T. Cvita\ifmmode~\check{s}\else \v{s}\fi{}, and J.~M. Hutson,
  ``Three-body nonadditive forces between spin-polarized alkali-metal atoms,''
  {\em Phys. Rev. A}, vol.~67,  054702, 2003.

\bibitem{Naidon2011tee}
P.~Naidon and M.~Ueda, ``The Efimov effect in lithium 6,'' {\em C.R. Physique},
  vol.~12,  13, 2011.

\bibitem{Hammer2007erc}
H.-W. Hammer, T.~A. L\"{a}hde, and L.~Platter, ``Effective-range corrections to
  three-body recombination for atoms with large scattering length,'' {\em Phys.
  Rev. A}, vol.~75, 032715, 2007.

\bibitem{Pricoupenko2010cef}
L.~Pricoupenko, ``Crossover in the Efimov spectrum,'' {\em Phys. Rev. A},
  vol.~82,  043633, 2010.

\bibitem{Wang2010utb}
Y.~Wang, J.~P. D'Incao, and B.~D. Esry, ``Ultracold three-body collisions near
  Feshbach resonances,'' {\em Phys. Rev. A}, vol.~83,  042710, 2011.

\bibitem{Thogersen2008upo}
M.~Th{\o}gersen, D.~V. Fedorov, and A.~S. Jensen, ``Universal properties of
  Efimov physics beyond the scattering length,'' {\em Phys. Rev. A}, vol.~78,
   020501(R), 2008.

\bibitem{Chen2011ttb}
C.~Ji, D.~R. Phillips, and L.~Platter, ``The three-boson system at
  next-to-leading order in the pionless EFT,'' {\em arXiv:1106.3837}, 2011.

\bibitem{Williams2009efa}
J.~R. Williams, E.~L. Hazlett, J.~H. Huckans, R.~W. Stites, Y.~Zhang, and K.~M.
  O'Hara, ``Evidence for an excited-state Efimov trimer in a three-component
  fermigas,'' {\em Phys. Rev. Lett.}, vol.~103,  130404, 2009.

\bibitem{Amado1973tin}
R.~D. Amado and F.~C. Greenwood, ``There is no Efimov effect for four or more
  particles,'' {\em Phys. Rev. D}, vol.~7,  2517, 1973.

\bibitem{Adhikari1981fbe}
S.~K. Adhikari and A.~C. Fonseca, ``Four-body Efimov effect in a
  Born-Oppenheimer model,'' {\em Phys. Rev. D}, vol.~24, 416--425,
  1981.

\bibitem{Naus1987tee}
H.~W.~L. Naus and J.~A. Tjon, ``The Efimov effect in a four-body system,'' {\em
  Few-Body Systems}, vol.~2, 121--126, 1987.

\bibitem{Sorensen2002ctb}
O.~S\o{}rensen, D.~V. Fedorov, and A.~S. Jensen, ``Correlated trapped bosons
  and the many-body Efimov effect,'' {\em Phys. Rev. Lett.}, vol.~89,
   173002, 2002.

\bibitem{Platter2004fbs}
L.~Platter, H.-W. Hammer, and U.-G. Mei\ss{}ner, ``Four-boson system with
  short-range interactions,'' {\em Phys. Rev. A}, vol.~70,  052101, 2004.

\bibitem{Yamashita2006fbs}
M.~T. Yamashita, L.~Tomio, A.~Delfino, and T.~Frederico, ``Four-boson scale
  near a Feshbach resonance,'' {\em Europhys. Lett.}, vol.~75,  555--561, 2006.

\bibitem{Hanna2006eas}
G.~J. Hanna and D.~Blume, ``Energetics and structural properties of
  three-dimensional bosonic clusters near threshold,'' {\em Phys. Rev. A},
  vol.~74,  063604, 2006.

\bibitem{Thogersen2008nbe}
M.~Th{\o}gersen, D.~V. Fedorov, and A.~S. Jensen, ``N-body Efimov states of
  trapped bosons,'' {\em Europhys. Lett.}, vol.~83,  30012, 2008.

\bibitem{Castin2010fbe}
Y.~Castin, C.~Mora, and L.~Pricoupenko, ``Four-body Efimov effect for three
  fermions and a lighter particle,'' {\em Phys. Rev. Lett.}, vol.~105,
   223201, 2010.

\bibitem{Hammer2007upo}
H.-W. Hammer and L.~Platter, ``Universal properties of the four-body system
  with large scattering length,'' {\em Eur. Phys. J. A}, vol.~32, 113--120,
  2007.

\bibitem{Vonstecher2009sou}
J.~{von Stecher}, J.~P. D'Incao, and C.~H. Greene, ``Signatures of universal
  four-body phenomena and their relation to the Efimov effect,'' {\em Nature
  Phys.}, vol.~5, 417--421, 2009.

\bibitem{Dincao2009ufb}
J.~P. D'Incao, J.~{von Stecher}, and C.~H. Greene, ``Universal four-boson
  states in ultracold molecular gases: Resonant effects in dimer-dimer
  collisions,'' {\em Phys. Rev. Lett.}, vol.~103,  033004, 2009.

\bibitem{Mehta2009gtd}
N.~P. Mehta, S.~T. Rittenhouse, J.~P. D'Incao, J.~von Stecher, and C.~H.
  Greene, ``General theoretical description of $n$-body recombination,'' {\em
  Phys. Rev. Lett.}, vol.~103,  153201, 2009.

\bibitem{Vonstecher2010wbc}
J.~{von Stecher}, ``Weakly bound cluster states of Efimov character,'' {\em J.
  Phys. B: At. Mol. Opt. Phys.}, vol.~43,  101002, 2010.

\bibitem{vonStecher2011ufa}
J.~von Stecher, ``Universal five- and six-body droplets tied to an Efimov
  trimer,'' {\em arXiv:1106.2319}, 2011.

\bibitem{Bartenstein2005pdo}
M.~Bartenstein, A.~Altmeyer, S.~Riedl, R.~Geursen, S.~Jochim, C.~Chin,
  J.~{Hecker Denschlag}, R.~Grimm, A.~Simoni, E.~Tiesinga, C.~J. Williams, and
  P.~S. Julienne, ``Precise determination of $^6$Li cold collision parameters
  by radio-frequency spectroscopy on weakly bound molecules,'' {\em Phys. Rev.
  Lett.}, vol.~94,  103201, 2005.

\bibitem{Lompe2010rfa}
T.~Lompe, T.~B. Ottenstein, F.~Servane, A.~N. Wenz, G.~Z\"urn, and S.~Jochim,
  ``Radio-frequency association of Efimov trimers,'' {\em Science}, vol.~330,
   940, 2010.

\bibitem{Dincao2006eto}
J.~P. D'Incao and B.~D. Esry, ``Enhancing the observability of the Efimov
  effect in ultracold atomic gas mixtures,'' {\em Phys. Rev. A}, vol.~73,
   030703(R), 2006.

\bibitem{Dincao2006mdo}
J.~P. D'Incao and B.~D. Esry, ``Mass dependence of ultracold three-body
  collision rates,'' {\em Phys. Rev. A}, vol.~73,  030702, 2006.

\bibitem{Nishida2008ufg}
Y.~Nishida and S.~Tan, ``Universal fermigases in mixed dimensions,'' {\em
  Phys. Rev. Lett.}, vol.~101,  170401, 2008.

\bibitem{Nishida2009cie}
Y.~Nishida and S.~Tan, ``Confinement-induced Efimov resonances in Fermi-Fermi
  mixtures,'' {\em Phys. Rev. A}, vol.~79,  060701(R), 2009.

\bibitem{Levinsen2009ads}
J.~Levinsen, T.~G. Tiecke, J.~T.~M. Walraven, and D.~S. Petrov, ``Atom-dimer
  scattering and long-lived trimers in fermionic mixtures,'' {\em Phys. Rev.
  Lett.}, vol.~103,  153202, 2009.

\bibitem{Wang2011eef}
Y.~Wang, J.~P. D'Incao, and C.~H. Greene, ``Efimov effect for three interacting
  bosonic dipoles,'' {\em Phys. Rev. Lett.}, vol.~106,  233201, 2011.

\bibitem{Petrov2004wbm}
D.~S. Petrov, C.~Salomon, and G.~V. Shlyapnikov, ``Weakly bound molecules of
  fermionic atoms,'' {\em Phys. Rev. Lett.}, vol.~93,  090404, 2004.

\bibitem{Vonstecher2007sad}
J.~{von Stecher} and C.~H. Greene, ``Spectrum and dynamics of the BCS-BEC
  crossover from a few-body perspective,'' {\em Phys. Rev. Lett.}, vol.~99,
   090402, 2007.

\bibitem{Liu2009vef}
X.-J. Liu, H.~Hu, and P.~D. Drummond, ``Virial expansion for a strongly
  correlated fermigas,'' {\em Phys. Rev. Lett.}, vol.~102,  160401, 2009.

\bibitem{Tan2009eoa}
S.~Tan, ``Energetics of a strongly correlated Fermi gas,'' {\em Ann. Phys.},
  vol.~323, 2952--2970, 2008.

\bibitem{Braaten2009htt}
E.~Braaten, ``How the tail wags the dog in ultracold atomic gases,'' {\em
  Physics}, vol.~2,  9, 2009.

\bibitem{Daley2009atb}
A.~J. Daley, J.~M. Taylor, S.~Diehl, M.~Baranov, and P.~Zoller, ``Atomic
  three-body loss as a dynamical three-body interaction,'' {\em Phys. Rev.
  Lett.}, vol.~102,  040402, 2009.

\end{thebibliography}


\end{document}